\RequirePackage{ifpdf}
\ifpdf 
\documentclass[pdftex]{sigma}
\else
\documentclass{sigma}
\fi

\usepackage[all]{xy}
\numberwithin{equation}{section}

\begin{document}
\allowdisplaybreaks

\renewcommand{\PaperNumber}{020}

\FirstPageHeading

\ShortArticleName{Duality-Symmetric Approach to General Relativity
and Supergravity}

\ArticleName{Duality-Symmetric Approach\\ to General Relativity
and Supergravity}

\Author{Alexei J. NURMAGAMBETOV}
\AuthorNameForHeading{A.J. Nurmagambetov}

\Address{A.I. Akhiezer Institute for Theoretical Physics, NSC
``Kharkov Institute of Physics\\ and Technology'', 1
Akademicheskaya Str., Kharkiv,  61108 Ukraine}

\Email{\href{mailto:ajn@kipt.kharkov.ua}{ajn@kipt.kharkov.ua}}

\ArticleDates{Received October 19, 2005, in f\/inal form February
03, 2006; Published online February 15, 2006}

\Abstract{We review the application of a duality-symmetric
approach to gravity and supergravity with emphasizing benef\/its
and disadvantages of the formulation. Contents of these notes
includes: 1) Introduction with putting the accent on the role of
dual gravity within M-theory; 2) Dualization of gravity with a
cosmological constant in ${\rm D}=3$; 3) On-shell description of
dual gravity in ${\rm D} > 3$; 4) Construction of the
duality-symmetric action for General Relativity with/without
matter f\/ields; 5) On-shell description of dual gravity in
linearized approximation; 6) Brief summary of the paper.}


\Classification{83E15; 83E50; 53Z05}

\begin{flushright}
\begin{minipage}{9cm}
{\it\small To memory Vladimir G.~Zima and Anatoly I.~Pashnev, my
colleagues and friends, whom I had a privilege to know, and who I
now miss.}
\end{minipage}
\end{flushright}

\section{Introduction}

We are living in an epoch of {\tt Duality}. Starting at the
f\/irst half of the last century with Quantum Mechanics, solid
states and condensed matter physics, Duality was recognized at
early 70th as a very useful tool in studying high energy physics
of strong interactions. Later on, Duality was successfully applied
for developing non-perturbative methods in QCD. Other interesting
observations on Duality in supergravities and perturbative String
theory lent credence to Duality as a key ingredient of a Unif\/ied
Theory, whatever it would be.

A most promising way we could follow in our quest for a Unif\/ied
Theory is the Superstring theory \cite{gsw87,k99,p98}. However,
the initial success in the development of string theory after the
``First Superstring revolution'' revealed a drawback, when at the
end of 80th it was understood that the map of perturbative
Superstring theory looks like

\begin{figure}[h!]
\centerline{\includegraphics{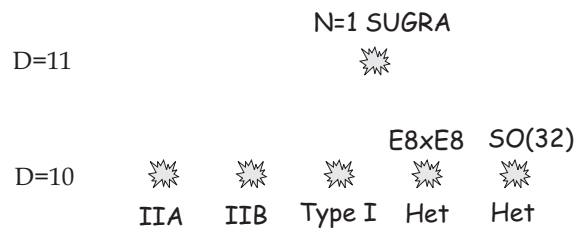}}
\caption{Strings' Theory map before the ``Second Superstring
Revolution''. }\label{Fi:Old}
\end{figure}

\noindent Clearly, this picture demonstrated a very strange and
unexpected feature of the String theory such as having f\/ive
consistent ten-dimensional Unif\/ied Theories instead of a naively
expected and desirable one Theory of Everything. Many questions
were also addressed to the role of eleven-dimensional supergravity
in the picture, since it was hardly related to any Strings, and
its interpretation within Superstring theory was obscured. These
and other related problems of String theory were resolved with
discovering dif\/ferent kinds of Dualities between String theories
in weak and strong coupling constant regimes \cite{ht95}. As a
result, the map of String theories was modif\/ied to

\begin{figure}[h!]
\centerline{\includegraphics{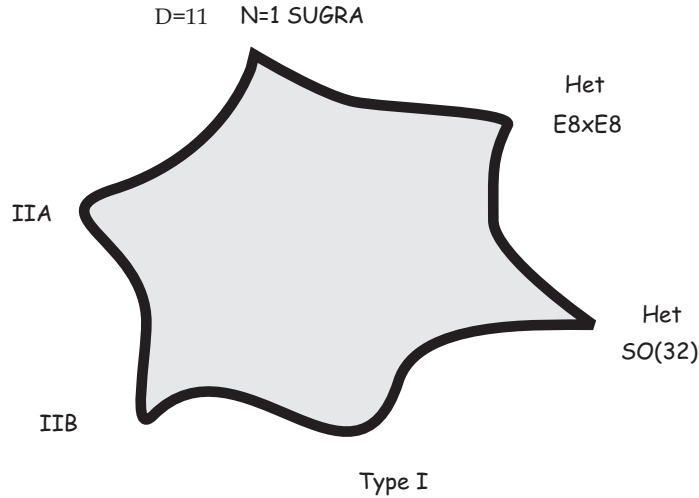}}
\caption{M-theory map.}
\end{figure}

\noindent The heart of the non-perturbative unif\/ication of
String theories was christened M-theory. During the last decade
M-theory was under the active studying, but one should ascertain
that our knowledge of M-theory structure is still far from
complete. Curiously enough, we had more strong evidence on
conjectured time ago quantum symmetries of M-theory such as
U-duality~\cite{op99}, having for the time being no classical
M-theory description which would contain classical U-duality as a
symmetry group. We can also discuss a Landscape of all possible
vacua of M-theory~\cite{s03,d04}, bearing in mind at the same time
a Landscepticism to this approach \cite{b04}. We can f\/ind
various scenario of appearing the de Sitter phase in M-theory
Cosmology \cite{linde05}, having nowadays no good mechanism of
supersymmetry breaking and a solution to the cosmological constant
problem. We have known the spectrum of free higher-spin f\/ields
in String theory, but it is not clear to date how to extract from
M-theory a detailed information on higher-spin f\/ields
interactions. One can continue this list of potential problems of
M-theory, and most of these problems are indications that little
is known on a microscopic formulation of M-theory.

The role of eleven-dimensional supergravity \cite{cjs78} in
M-theory picture becomes transparent: it is the low-energy limit
of M-theory. In its turn, ten-dimensional supergravities \cite{ss}
correspond to the low-energy limits of dif\/ferent superstrings
when strings tension goes to inf\/inity. Many initial conjectures
on dualities between string theories and M-theory were
successfully proved for corresponding supergravities. And vice
versa, many substantial fragments of Dualities in supergravities
were regained for Strings.

Though Supergravity has been approved as a very useful tool in
studying M-theory, such an approach is apparently restrictive. The
f\/ield content of (ungauged) ${\rm D}=10$ supergravities consists
only of massless f\/ields which form the bottom of the whole
spectrum of superstrings' excitations. The total number of
superstring modes is inf\/inite, and other f\/ields of Superstring
spectrum are massive, with masses of an order of the Plank mass.
Clearly, these f\/ields are too heavy to include it in
Supergravity as a low-energy ef\/fective theory constructed out of
light f\/ields. However, in the opposite to the low energy limit
the string tension goes to zero, all the String modes become
massless, and we are going to tell on the high energy limit of
String theory, where the full hidden symmetry of String theory is
restored. (To make the discussion on various limits of String
theory more transparent to the reader we add the corresponding
Appendix.) An ef\/fective f\/ield theory which will describe the
high energy limit of String theory and will manifestly be
invariant under the full symmetry of String theory has to deal
with inf\/inite set of massless higher-spin f\/ields, and shall be
a dual to Supergravity theory \cite{gross88}. The strong coupling
limit of String theory could be obtained then via the hidden
symmetry breaking upon coming from ultra-high energies to the
Plank energy. Hence, f\/iguring out the full hidden symmetry
structure will supply us with an important information on the
ef\/fective action of String theory in the strong coupling regime.

So far we did not specify the String theory we will consider in
the strong coupling regime. However, the strong coupling limit of
type IIA superstring theory is precisely M-theory
\cite{towns95,witten95}, so following the way of the reasoning in
the above we have to conclude that a symmetry of M-theory shall be
big enough to accommodate inf\/inite number of f\/ields: it {\tt
has to be} inf\/inite-dimensional!

An inf\/inite-dimensional algebraic structure is common for
Strings. A f\/irst quantized (super)string theory is managed by
the inf\/inite-dimensional Virasoro algebra \cite{vir70,fv71},
which is a~particular example of the Kac--Moody algebras
\cite{k83}. The inf\/inite-dimensional structure of an M-theory
symmetry should have an impact on the low-energy ef\/fective
action. So, we could pose a natural question: how would we observe
fragments of this inf\/inite-dimensional symmetry inside of ${ \rm
D}=11$ supergravity?

A way of producing Kac--Moody structures in Supergravities is
known: they come from the toroidal dimensional reduction of ${\rm
D}=10/11$ Supergravities. To be precise, the following chain of
$\mathrm{E}_n$ groups ($\mathrm{E}_1 \sim R^{+}$,
$\mathrm{E}_2=\mathrm{A}_1$, $\mathrm{E}_3 =\mathrm{A}_2 \oplus
\mathrm{A}_1$, $\mathrm{E}_4 = \mathrm{A}_4$, $\mathrm{E}_5 =
\mathrm{D}_5$; see e.g.\ \cite{op99})
\[ \mathrm{E}_1 \longrightarrow \mathrm{E}_2 \longrightarrow \mathrm{E}_3
\longrightarrow \mathrm{E}_4 \longrightarrow \mathrm{E}_5
\longrightarrow \mathrm{E}_6 \longrightarrow \mathrm{E}_7
\longrightarrow \mathrm{E}_8
\] appears under dimensional reduction
of ${\rm D}=11$ supergravity on
\[ T_1 \longrightarrow T_2 \longrightarrow T_3 \longrightarrow T_4
\longrightarrow T_5 \longrightarrow T_6 \longrightarrow T_7
\longrightarrow T_8
\]
 and these global symmetry groups are in
general hidden.  To recover such a hidden symmetry structure one
should dualize higher-rank f\/ields, with the rank $r>(D-n)/2$,
that appear upon the reduction on a n-dimensional torus $T_n$
\cite{cj79,cjlp98I,cjlp98II}.

So far we did not succeed in our quest of a Kac--Moody structure.
However, let us take a~conjecture that doing the reduction to
${\rm D}=2$ \cite{julia82,julia84,nw89}, ${\rm D}=1$
\cite{julia84, nicolai92} and, f\/inally, to ${\rm
D}=0$~\cite{moore93} we will further continue the
$\mathrm{E}_n$-series like
\[
\cdots  \longrightarrow \mathrm{E}_9 \longrightarrow
\mathrm{E}_{10} \longrightarrow \mathrm{E}_{11} .
\]
 Once we accept this, we get
what we need, because, algebraically, $\mathrm{E}_9$ is the
(unique possible) af\/f\/ine extension of $\mathrm{E}_8$, and a
non-trivial af\/f\/ine extension of any Lie algebra corresponds to
a Kac--Moody algebra. The structure of $\mathrm{E}_{10}$,
$\mathrm{E}_{11}$ is, of course, more complicated, but anyway this
is an extension of the Kac--Moody structure.

Remarkably, it was proved that the conjecture does work, at least
when one reduces ${\rm D}=11$ Supergravity to two and to one
dimensions, and indeed, $\mathrm{E}_9$ and $\mathrm{E}_{10}$
global symmetry algebras appear \cite{nw89,miz98}.  Hence, it is
very likely  that we can expect a still elusive $\mathrm{E}_{11}$
to also appear in the end.

We have sketched out a scheme of appearance of the Kac--Moody
algebras inside of Supergravity, but in a reduced theory. It is
not a priori obvious that there exists a way to relate algebraic
structures of a reduced theory with that of the unreduced theory.
But it turned out that it is really possible to identify the
global symmetry groups of the dimensionally reduced on $T_n$
supergravities to the hidden symmetry group of M-theory
\cite{wn00}. Some observations in favor of such an
identif\/ication were made in mid 80th  \cite{wn85,wn86,n87}, and
following them it was demonstrated \cite{kns00} that the
``exceptional geometry" of the ${\rm D}=3$ maximal Supergravity
admits a re-formulation of ${\rm D}=11$ Supergravity in a
$\mathrm{E}_8$ invariant way. Put it another way, some of the
objects which appear upon the reduction do naturally appear in the
unreduced theory.

More evidence in favor of $\mathrm{E}_{11}$ as a hidden symmetry
of M-theory was found in \cite{w00,w01,sw01,sw02,w02}. There it
was considered a non-linear realization of gravity and of the
bosonic sector of maximal supergravities which is based on a
generalization of Borisov--Ogievetsky formalism \cite{bo74,o73}.
As well as ${\rm D}=4$ gravity \cite{bo74,o73}, ${\rm D}=11$
supergravity can be realized as to be invariant under an
inf\/inite-dimensional group, which is the closure of a
$\mathrm{G}_{11}$ invariance group \cite{w00} and the conformal
group $\mathrm{SO(2,10)}$. The group $\mathrm{G}_{11}$ is formed
out of the af\/f\/ine group $\mathrm{IGL(11)}$ extended with the
subgroup which is generated by antisymmetric type generators with
three and six indices. These new generators correspond to $A_{c_1
c_2 c_3}$ gauge f\/ield of ${\rm D}=11$ supergravity and to its
dual partner $A_{c_1 c_2 \cdots c_6}$, and form a subalgebra.

Being f\/inite-dimensional, $\mathrm{G}_{11}$ is apparently not a
Kac--Moody algebra, and its coset representatives (see \cite{w00})
are not members of a Borel subgroup of a larger group as it should
be for the coset formulation of the dimensionally reduced ${\rm
D}=11$ supergravity \cite{cjlp98I,cjlp98II}. A way to overcome
this obstacle is to invoke more than just antisymmetric gauge
f\/ields duality in the picture and to enlarge $\mathrm{G}_{11}$
with a generator corresponding to a dual to graviton f\/ield
\cite{w01}. Then it turns out that simple roots of a Kac--Moody
algebra we seek are those of simultaneously sharing
$\mathrm{SL(11)}$

\vspace{0.3cm}
\begin{figure}[h!]
\centerline{\includegraphics{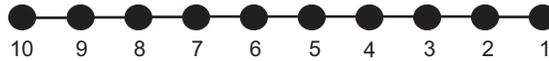}} \caption{Dynkin
diagram of $\mathrm{SL(11)}$.}
\end{figure}

\noindent and $\mathrm{E}_8$ subalgebras

\begin{figure}[h!]
\centerline{\includegraphics{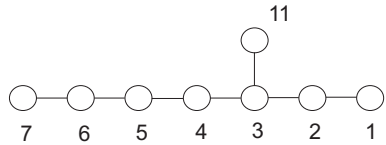}} \caption{Dynkin
diagram of $\mathrm{E}_8$.}
\end{figure}

Gluing two Dynkin diagrams together leads to desired
$\mathrm{E}_{11}=\mathrm{E}_8^{+++}$

\vspace{0.1cm}
\begin{figure}[h!]
\centerline{\includegraphics{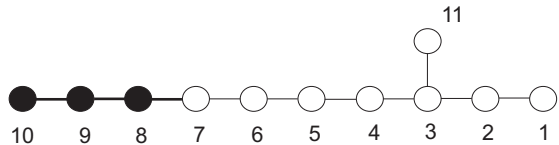}} \caption{Dynkin
diagram of $\mathrm{E}_{11}$.}
\end{figure}

\noindent which is nothing but the extension of the $\mathrm{E}_8$
Dynkin diagram with three additional nodes, called the
very-extension of $\mathrm{E}_8$ \cite{w02,gow02}.

It has to be emphasized that a dual to graviton f\/ield was a
missing point which should be recovered for the following reasons.
From the algebraic point of view, having new generator of the dual
to graviton f\/ield in a reduced $\mathrm{E}_{11}$ algebra is
crucial for recovering the correct algebra for a non-linear
realization of type IIA supergravity \cite{w02}. This algebra
includes generators of f\/ields entering type IIA supergravity
multiplet together with generators which correspond to their Hodge
dual partners. Hence, a non-linear realization based on this
algebra results in a~duality-symmetric formulation of type IIA
supergravity when f\/ields of the supergravity multiplet and their
duals enter the formulation on an equal footing and from the very
beginning. In its turn, the structure of $\mathrm{G}_{11}$
dictates the duality-symmetric structure of ${\rm D}=11$
supergravity. Such a duality-symmetric formulation of the
antisymmetric gauge f\/ield sector was proposed in \cite{bbs98} to
couple a dynamical M5-brane \cite{blnpst97,apps97} to ${\rm D}=11$
supergravity. However, having just duality-symmetric structure of
the gauge sector in ${\rm D}=11$ is not enough to recover
duality-symmetric structure of type IIA supergravity after
reduction \cite{bns}. It was straightforward once a dual to
graviton f\/ield would be taken into account \cite{ajnjetp04},
that we are further going to demonstrate.

Any duality-symmetric formulation double physical, i.e. on-shell,
degrees of freedom  as compared to a standard approach. This
mismatch is removed by imposing relations between dual partners,
the so-called duality relations. A minimal, but not convenient for
further possible applications, way to take duality relations into
account is to impose them by hand as an additional part to
equations of motion (see e.g.~\cite{bkorvp01}). Following this
way, we have to bear in mind that further modif\/ications of
equations of motion, such as coupling to external sources,
introducing a non-trivial self-interaction of f\/ields, taking
into account quantum corrections etc., will require an appropriate
change of the duality relations which would be hard in general to
guess. More conveniently, but technically more complicated, is to
f\/igure out a way to derive duality relations from an action as
equations of motion (see~\cite{s02} for a review). There are many
roads to this end; here we will mainly follow
\cite{pst95plb,pst95prd,pst97prd}, reviewing in brief other
approaches.

To summarize our Introduction, let us recall that the
duality-symmetric structure of ${\rm D}=11$ supergravity is very
natural on the way of dynamical realization of $\mathrm{E}_{11}$
conjecture, and if we require the hidden symmetry algebra to be a
Kac--Moody-type algebra, we have to f\/igure out a~way of
dualization of gravity.

The rest of the paper is organized as follows: in the
following section we consider a warm-up exercise of dualization of
three-dimensional gravity with a cosmological constant. Though
this example cannot be considered as a general case, it
nevertheless demonstrates some common features of the dualization
which will be helpful in further generalization of the dual
description to higher dimensions. In Section~3 we construct the
on-shell dual gravity in the space-time of dimensions higher than
three. Then, in Section~4 we uplift our formulation at the
of\/f-shell level. A brief review of the construction in the
linearized approximation is given in Section~5, and our
conclusions are summarized in the last section. Remarks on the
construction, discussion on formulations we use and examples of
application are adduced throughout the paper. Our notation and
conventions, as well as an informative summary on dif\/ferent
limits of String theory, are collected in two Appendices.

\section[Dualization of D = 3 gravity]{Dualization of $\boldsymbol{{\rm D}=3}$ gravity}

We are getting started with the action of ${\rm D}=3$ gravity with
a cosmological constant $\lambda$
\begin{gather}\label{GR3}
S=\int  d^3 x \, \sqrt{g}(R+\lambda),
\end{gather}
and the inclusion of non-zero cosmological constant into the
action is important for the subsequent dualization of the model.
It will also be convenient for this purpose to present the
originally second order action (\ref{GR3}) in the form of the
Hilbert--Palatini f\/irst order action
\begin{gather}\label{GR3HP}
S=\int  d^3 x \, \left[ \epsilon^{mnk} e_{ka} \left(\frac{1}{2}
{R_{mn}}^{bc} \epsilon_{abc}\right)+\lambda \det e^a_m \right].
\end{gather}

In the f\/irst order formulation a dreibein $e^a_m$ and
$\mathrm{SO(1,2)}$ spin connection $\omega_m^{ab}$ are supposed to
be independent variables. The curvature tensor depends only on the
connection
\begin{gather*}
{R^{ab}}_{mn}=2\partial_{[m} \omega^{ab}_{n]}+2\omega^{ac}_{[m}
\omega^{db}_{n]}\eta_{cd}\equiv \partial_{m}
\omega^{ab}_{n}-\partial_{n} \omega^{ab}_{m}+\big(\omega^{ac}_{m}
\omega^{db}_{n}-\omega^{ac}_{n} \omega^{db}_{m}\big)\eta_{cd},
\end{gather*}
where $\eta_{ab}$ is the Minkowski metric tensor with the mostly
minus signature (see Appendix~A for the notation). As a result,
the Einstein equation that comes from the f\/irst order
action~(\ref{GR3HP}) is of the f\/irst order in the connection. To
arrive at the standard second order equation of motion of a~spin
two bosonic f\/ield, one should violate the initial independence
of variables and to resolve an algebraic relation between the
dreibein and the spin connection. This relation follows
from~(\ref{GR3HP}) when one varies the action over the spin
connection as an independent variable. Once the connection is
expressed in terms of the dreibeins the standard second order
Einstein equation is recovered. Therefore, the dreibeins are the
true dynamical variables of the theory, while the spin connection
plays the role of an additional but not of a fundamental variable.

However, it is possible to replace vielbeins with other
fundamental variables. Let us note to this end that the form of
(\ref{GR3HP}) suggests the def\/inition of a new connection
\begin{gather*}
A^a_m=\frac{1}{2} \epsilon^{abc} \omega_{mbc}
\end{gather*}
with the same as the dreibeins index structure. In terms of the
new variables the curvature tensor acquires the form of a
Yang--Mills f\/ield strength \cite{peldan}
\begin{gather}\label{D3F}
F^a_{mn}=2\partial_{[m} A^a_{n]}+\epsilon^{abc} A_{mb} A_{nc}.
\end{gather}
Then the Hilbert--Palatini action (\ref{GR3HP}) becomes
\begin{gather}\label{D3grac}
S=\int  d^3 x \left[ \epsilon^{mnk} e_{ka} F_{mn}^a+\lambda \det
e^a_m \right],
\end{gather}
and it is easy to see that the true equation of motion following
from (\ref{D3grac}) upon varying over the dreibein is
\begin{gather}\label{D3eom}
*F^{ma}+\lambda e e^{ma}=0.
\end{gather}
We have denoted $e=\det e^a_m$ in the above and have used the
following def\/inition of the dual, with respect to the curved
indices, tensor
\begin{gather*}
*F^{ma}=\epsilon^{mnk} F^a_{nk}.
\end{gather*}
The other equation which can be read of\/f (\ref{D3grac}) is
\begin{gather}\label{T3eom}
{\mathcal D}_m (\epsilon^{mnk} e_{ka})=0.
\end{gather}
Here ${\mathcal D}_m$ is the covariant with respect to $A^a_m$
derivative, and (\ref{T3eom}) is the algebraic relation which
expresses the initially independent variables $A^a_m$ through
$e^a_m$.

As for the standard f\/irst order formulation of gravity in
$(e^a_m,\omega^{ab}_m)$ variables the integrability condition to
equations (\ref{T3eom}), (\ref{D3eom}), viz.
\begin{gather}
\mathcal{D}_m \mathcal{D}_n (\epsilon^{mnk} e_{ka})=0,\nonumber
\\
\label{integroF3} \mathcal{D}_m (\ast F^{ma}+\lambda e e^ma)=0
\end{gather}
results in the f\/irst-type (for torsion) and in the second-type
(for curvature) Bianchi identities respectively. We remind that in
the former set of variables these Bianchi identities are
${R^{a}}_{[nkl]}=0$ and $\mathcal{D}_{[m} {R^{ab}}_{nk]}=0$. Note
also that taking into account (\ref{T3eom}) the integrability
condition (\ref{integroF3}) can be rewritten as
\begin{gather}\label{BIF3}
\mathcal{D}_m (\epsilon^{mnk} F^a_{nk})=0,
\end{gather}
which is nothing but the Bianchi identity for the Yang--Mills
f\/ield strength (\ref{D3F}). Strictly spea\-king, the Bianchi
identity (\ref{BIF3}) has nothing to do with the integrability
condition (\ref{integroF3}), since it directly follows from the
def\/inition (\ref{D3F}); rather the integrability condition
(\ref{integroF3}) is satisf\/ied due to the zero torsion equation
(\ref{T3eom}) and the Bianchi identity (\ref{BIF3}).

So far we considered $e^a_m$ as the fundamental variables.
However, we can also think about $A^a_m$ as the other fundamental
variables. To go from one to another we shall use
equation~(\ref{D3eom}).

Taking the determinant of (\ref{D3eom}) leads to
\begin{gather}\label{D3detEOM}
-\det (*F^{ma})=\lambda^3 e^2,
\end{gather}
after that one can resolve (\ref{D3eom}) as
\begin{gather}\label{D3eomR}
e^{ma}=\mp \frac{1}{\lambda
\sqrt{-\frac{1}{\lambda^3}\det(*F^{ma})}} *F^{ma}.
\end{gather}
In ef\/fect, we have completely determined the dreibein in terms
of the dual curvature or, equivalently, in terms of the new
connection $A^a_m$. Substituting this solution to the Lagrangian
back results in
\begin{gather}\label{D3dgrac}
S=\mp 2 \,\mathrm{sign} (\lambda) \int d^3 x
\sqrt{-\frac{1}{\lambda} \det (*F^{ma})}
\end{gather}
that is the action of the model in terms of $A^a_m$.

Let us now compare the equation of motion following from
(\ref{D3dgrac}) with equations (\ref{D3eom}) and~(\ref{T3eom}).
Varying (\ref{D3dgrac}) with respect to $A^a_m$ we get the
Yang--Mills-type equation of motion (see~\cite{peldan} for
details)
\begin{gather}\label{FD3eom}
{\mathcal D}_m \big(\sqrt{g} F^{mn}_a\big)=0
\end{gather}
on a curved manifold endowed with the metric depending on $A^a_m$.
As it follows from (\ref{D3detEOM})
\begin{gather*}
\sqrt{g}=\sqrt{-\frac{1}{\lambda^3}\det (*F^{ma})}.
\end{gather*}
In addition we have to complete (\ref{FD3eom}) with the Bianchi
identity (\ref{BIF3}) which follows from the def\/inition of the
new curvature (\ref{D3F}).

One could note that equation (\ref{FD3eom}) is nothing but
equation (\ref{T3eom}) written down in terms of the new variables
when (\ref{D3eom}) is taken into account. In its turn the Bianchi
identity (\ref{BIF3}) follows from~(\ref{integroF3}) as the
integrability condition to the relation (\ref{D3eom}) that is
algebraic in this picture. Put it differently an
algebraic relation that comes from the action in standard
variables becomes the equation of motion of the model in new
variables. This phenomenon is a remnant of the dualization, when
the Bianchi identities (algebraic relations) of gauge f\/ields
become equations of motion of the dual gauge f\/ields. Hence, one
can say that (\ref{D3dgrac}) is the action for gravity in dual
variables, or, for short, the dual gravity.

What we can learn from this example? We have observed that the
problem is completely resolved in ${\rm D}=3$ case. However, it is
only resolved for gravity with a cosmological constant, and the
dual Lagrangian is not well def\/ined in the zero cosmological
constant limit. Furthermore, the case under consideration cannot
be treated as a general case, and the above mentioned way of the
dualization is not a general scheme, since gravity in ${\rm D}=3$
possesses the features which cannot be encountered for
gravitational theory in other space-time dimensions. For instance,
it is the very case where the dualized over SO(1,2) indices
connection has the same index structure as the dreibein, and we
have used this fact to completely get rid of the dreibeins as
fundamental dynamical variables in favor of the dual connection.
It would be pertinent to note that using a connection as a
fundamental variable is one of the features of the Loop Quantum
Gravity (see~\cite{npz05} for a recent review). Also, ${\rm D}=3$
gravity does not have dynamical degrees of freedom~
\cite{djh84,dj84,dj82}, and is interesting rather as a topological
exactly (classically and quantum) solvable theory~\cite{w88}.

Taking into account the above, perspectives to generalize such a
dual description to higher dimensions do not seem to be so direct.
Starting with ${\rm D}=4$ gravity with a cosmological constant we
will not have so simple form of equations of motion that would
suggest a way of resolving vielbeins in terms of the dual
curvature/dual connection. For example, applying the determinant
to both sides of the four-dimensional Einstein equation with a
cosmological constant does not lead to the dual theory. It results
in the Eddington--Schr\"odinger formulation of
gravi\-ty~\cite{schro,ft85} which is classically equivalent to the
Einstein formulation (see \cite{dg98,dn82} for other treatments of
the problem). But even if we were succeed in dualization of
higher-dimensional gravity with a cosmological constant it would
not be a resolution to the problem in the M-theory context, since
there is no place for a cosmological constant in ${\rm D}=11$
supergravity \cite{deserseminara}.

Finally, we have to emphasize an important point: because of the
square root the dual Lagrangian resembles the Nambu--Goto action
for strings and branes, it is non-polynomial and it has inf\/inite
number of terms with inf\/inite number of derivatives in the small
cosmological constant limit $\lambda \rightarrow 0$ supporting
with the weak f\/ield approximation. In this limit the gravity can
be approximated with a small perturbation around Minkowski f\/lat
space (it does not contradict the results of~\cite{cd80} since the
cosmological constant tends to zero in the limit), and passing to
a new curvature through $\ast F^{ma}=\ast F^{mn}e^a_n$ we get
f\/irst
\begin{gather*}
\det {(\ast F^{ma})}=\det{e^a_m} \det{(\ast F^{mn})}.
\end{gather*}
Second, we approximate the new curvature as
\begin{gather}\label{FD3app}
\ast F^{mn}=\Lambda \eta^{mn}+\ast R^{mn},
\end{gather}
where $\eta^{mn}$ is the inverse to the Minkowski f\/lat metric
tensor, and we have f\/ixed for def\/initeness $\lambda < 0$,
$\Lambda=|\lambda|$. Also we would like to f\/ix the ``minus''
sign on the r.h.s.\ of (\ref{D3eomR}), so under these assumptions
the Lagrangian for the dual gravity becomes
\begin{gather*}
\mathcal{L}=2\sqrt{\frac{1}{\Lambda} \det{e^a_m} \det{(\ast
F^{mn})}}.
\end{gather*}
Substituting (\ref{FD3app}) into the above, we arrive at
\begin{gather*}
\mathcal{L}=2\Lambda\sqrt{\det{e^a_m}}\sqrt{\det{(\delta_m^n+\frac{1}{\Lambda}
\ast R^{n}_m)}},
\end{gather*}
that after taking into account $\det{M}=\exp{(\mathrm{Tr} \ln M)}$
results in
\begin{gather*}
\mathcal{L}=\sqrt{\det{e^a_m}} \left[ 2\Lambda+\ast
{R^m}_m-\frac{1}{2\Lambda}\left( \ast R^{mn} \ast
R_{mn}-\frac{1}{2}(\ast {R^m}_m)^2 \right)+\cdots \right].
\end{gather*}

This reminds the situation which one encounters in
{\tt{non-local}} theories. However, in our example this
non-polynomiality appeared because of the choice of dual variables
and, as in the case of the Nambu--Goto action, it is harmless. At
the same time this observation suggests that in a general case
proper non-locality could appear as a by-product of dualization.

\section[On-shell dual gravity in ${\rm D} > 3$: Bianchi vs. Einstein]{On-shell dual gravity in
$\boldsymbol{\mathrm{D}>3}$: Bianchi vs.\ Einstein}

After an examination of three-dimensional gravity, let us try to
f\/igure out a way of proceeding in $\mathrm{D>3}$ case. What we
shall f\/ind f\/irst is a convenient representation of the
Einstein equation in a way that allows us to present it as the
Bianchi identity for a dual f\/ield. It turns out to be convenient
for this purpose to write down the Einstein equation in a form
which is similar to the dynamics of Maxwell theory with a source.

\subsection{Maxwell-like representation of the Einstein equation}

To arrive at such a representation let us get started with the
Hilbert--Palatini action of $\mathrm{D}$-di\-mensional gravity
\begin{gather}\label{Dgrac}
S_{HP}=\int_{{\mathcal M}^D}  \hat{R}^{\hat{a}_1 \hat{a}_2} \wedge
\hat{\Sigma}_{\hat{a}_1 \hat{a}_2},
\end{gather}
where
\begin{gather*}
\hat{R}^{\hat{a} \hat{b}}=d\hat{\omega}^{\hat{a}
\hat{b}}-{\hat{\omega}^{\hat a}}_{~\hat c} \wedge
\hat{\omega}^{\hat{c} \hat{b}}
\end{gather*}
is the curvature two-form expressed via $SO(1,D-1)$ connection
$\hat{\omega}^{\hat{a}\hat{b}}$ and
\begin{gather*}
\hat{\Sigma}_{{\hat a}_1\cdots {\hat a}_n}=\frac{1}{
(D-n)!}\epsilon_{{\hat a}_1\cdots {\hat a}_{D}}{\hat e}^{{\hat
a}_{n+1}}\wedge\cdots\wedge {\hat e}^{{\hat a}_{D}}
\end{gather*}
is a $(\mathrm{D}-n)$-form constructed out of vielbeins ${\hat
e}^a$. Since we will use in what follows the reduction of the
action to $(\mathrm{D}-1)$-dimensions, we have reserved hats to
distinguish $\mathrm{D}$-dimensional indices and variables from
$(\mathrm{D}-1)$-dimensional quantities. This notation is generally
accepted in the Kaluza--Klein literature, and we will follow it.

Upon varying with respect to the vielbein and the connection we
get the Einstein equation of motion
\begin{gather}\label{Rem}
\hat{\Sigma}_{{\hat a}{\hat b}{\hat c}}\wedge \hat{R}^{{\hat
b}{\hat c}}=0,
\end{gather}
and an algebraic relation
\begin{gather*}
\hat{\Sigma}_{{\hat a}{\hat b}{\hat c}}\wedge \hat{T}^{\hat c}=0,
\end{gather*}
which sets the torsion 2-form $\hat{T}^{\hat a}=d\hat{e}^{\hat
a}-\hat{e}^{\hat b}\wedge{{\hat \omega}_{\hat b}}^{~{\hat a}}$ to
zero. The latter allows expressing the originally independent
connection $\hat{\omega}^{{\hat a}{\hat b}}$ through vielbeins and
their derivatives
\begin{gather}\label{om}
{\hat \omega}^{{\hat a}{\hat b}}({\hat
e})=\frac{1}{2}~\hat{e}^{\hat c}\left[ \hat{e}^{\hat m}_{\hat
c}\hat{e}^{{\hat n}{\hat a}}\partial_{[{\hat m}}\hat{e}_{{\hat
n}]}^{\hat b}-\hat{e}^{\hat m}_{\hat c}\hat{e}^{{\hat n}{\hat
b}}\partial_{[{\hat m}}\hat{e}_{{\hat n}]}^{\hat a}-\hat{e}^{{\hat
n}{\hat a}}\hat{e}^{{\hat s}{\hat b}}
\partial_{[{\hat n}}\hat{e}_{{\hat s}]{\hat c}} \right].
\end{gather}

After replacing the connection with (\ref{om}) the Einstein
equation is written down in the following form (see \cite{ajn04})
\begin{gather}\label{Rem3}
d(\hat{\ast} d\hat{e}_{\hat
a})=\hat{\ast}\hat{\tilde{J}}^{[1]}_{\hat a},
\end{gather}
where a one-form $\hat{\tilde{J}}^{[1]}_{\hat a}$ is
\begin{gather}\label{tilJ}
\hat{\tilde{J}}^{[1]}_{\hat
a}=(-)^{\frac{D(D-5)}{2}}\hat{J}^{[1]}_{\hat a}+\hat{\ast}d
\hat{S}^{[D-2]}_{\hat a},
\end{gather}
and
\begin{gather}\label{J}
\hat{J}^{[1]}_{\hat a}=\hat{\ast}\left[ \hat{\omega}^{{\hat
b}{\hat c}}({\hat e})\wedge d\hat{\Sigma}_{{\hat a}{\hat b}{\hat
c}}+(-)^{D-3}~{{\hat \omega}}^{\hat b}_{~{\hat d}}({\hat e})\wedge
\hat{\omega}^{{\hat d}{\hat c}}({\hat e})\wedge
\hat{\Sigma}_{{\hat a}{\hat b}{\hat c}} \right],
\\
\label{S} \hat{S}^{[D-2]}_{\hat a}=\hat{\ast} (\hat{e}^{\hat
b}\wedge \hat{e}_{\hat a})~ \hat{e}^{\hat m}_{\hat c}
\hat{e}^{\hat n}_{\hat b}\partial_{[\hat{m}} \hat{e}^{\hat
c}_{{\hat n}]}.
\end{gather}

\vspace{0.3cm} \noindent Two remarks are appropriate here.

\begin{remark}
Equation (\ref{Rem3}) looks like that of the Maxwell
electrodynamics with an electric-type source. This form of the
Einstein equation is not a new one; it has been pointed out in
literature for a long time (see e.g.\ \cite{thirring86,blau87}).
 Since the equation is written in terms of
dif\/ferential forms, it is manifestly invariant under general
coordinate transformations. But its manifest invariance under
local Lorentz rotations is lost. However, the right hand side of
(\ref{Rem3}) is a modif\/ied Landau--Lifshitz pseudo-tensor, so
the non-covariance on the left hand side is compensated by the
pseudo-tensor-type transformations on the right hand side. Hence,
equation (\ref{Rem3}) is covariant under the general coordinate
transformations and the Lorentz rotations in the tangent space.
\end{remark}

\begin{remark}
In the form of (\ref{Rem3}) the Einstein equation is closely
related to the so-called ``telepa\-rallel'' version of General
Relativity (see~\cite{pereira05} for a review).
\end{remark}

As a next step towards realization of our program we have to
rewrite the Einstein equation~(\ref{Rem3}) as the Bianchi identity
of a dual f\/ield. The Poincar\'e lemma claims that in a trivial
space-time topology setting one can present the conserved
``current'' form $\hat{\tilde{J}}^{[1]}_{\hat a}$ as the curl of
a~``pre-current'' form
\begin{gather}\label{tJ}
\hat{\ast}\hat{\tilde{J}}^{[1]}_{\hat
a}=d\hat{\ast}\hat{\tilde{G}}^{[2]}_{\hat a}.
\end{gather}
With account of it, the Einstein equation becomes $d (\cdots)=0$
which is the Bianchi identity for a dual f\/ield. That is exactly
what we need. Therefore, to make the f\/inal step from Einstein to
Bianchi, we have to resolve (\ref{tJ}) to f\/ind $\hat{\tilde
G}^{[2]}_{\hat a}$.

As to the explicit expression of the ``current'' (cf.\ equations
(\ref{tilJ}), (\ref{J}), (\ref{S})) it is not so simple to derive
the ``pre-current'' $\hat{\tilde{G}}^{[2]}_{\hat a}$ exactly, and
in a local form. Furthermore, there are no general arguments that
$\hat{\tilde{G}}^{[2]}_{\hat a}$ has always to be a local
expression. So it does not become a surprise that a general
procedure of resolving (\ref{tJ}) sacrif\/ices locality and leads
to local expressions for ``pre-currents'' only for ``currents'' of
a special type.

\subsection{Resolving via inverse d'Alembertian}

To write down equation (\ref{Rem3}) in terms of a (Hodge) dual
f\/ield one has to introduce the dual f\/ield strength
\begin{gather*}
\hat{F}_{\hat a}=\hat{\ast}d\hat{e}_{\hat a},
\end{gather*}
after that the Einstein equation recasts in
\begin{gather}\label{Rem4}
d\hat{F}_{\hat a}=\hat{\ast}\hat{\tilde{J}}^{[1]}_{\hat a}.
\end{gather}

The general solution to an equation
\begin{gather}\label{eq}
dK=\mathcal{J};\qquad \mathcal{J}\ne \mathcal{J} (K),
\end{gather}
which is presumably the case under consideration, is as follows
\begin{gather}\label{sol1}
K=\Box^{-1} {\boldsymbol \delta} \mathcal{J}+d\chi.
\end{gather}
Here ${\boldsymbol \delta}=(-)^{(D-p)p-1} \ast d \ast$ is the
$D$-dimensional co-derivative operator acting on a $p$-form,
$\Box=d{\boldsymbol \delta}+{\boldsymbol \delta} d$ is the
d'Alembertian and $\chi$ is a dif\/ferential form of the
appropriate rank. To verify~(\ref{sol1}) one should notice the
Hodge identity
\begin{gather*}
d\Box^{-1}{\boldsymbol \delta}+{\boldsymbol \delta}\Box^{-1}d=1,
\end{gather*}
which is easy to prove with taking into account
\begin{gather}\label{dtri}
[d,\Box]=0,\qquad [{\boldsymbol \delta},\Box]=0.
\end{gather}
These relations follow from the d'Alembertian def\/inition. Then,
for any form $\Omega$
\begin{gather}\label{mcJ}
\Omega=1\cdot \Omega=d\big(\Box^{-1}{\boldsymbol \delta}
\Omega\big)+{\boldsymbol \delta}\big(\Box^{-1} d \Omega\big),
\end{gather}
and the last term on the r.h.s.\ of (\ref{mcJ}) vanishes for a
closed form. Thus, the solution (\ref{sol1}) is the sum of a
particular solution to the non-homogeneous dif\/ferential equation
(\ref{eq}) and the general solution to the homogeneous equation
$dK=0$.

What is a restriction on this approach? If $\mathcal{J}$ is a
co-closed form, i.e.\ ${\boldsymbol \delta} \mathcal{J}=0$, or in
other words is a zero mode of the d'Alembertian, the way of the
reasoning apparently fails. Hence, the current form $\mathcal{J}$
has not to be co-closed.

But what we have to do if nevertheless the current is a co-closed
form? In such a case we have to apply a regularization. There are
many ways to this end, we will treat it as follows. Recall the
following integral representation for a variable $\rho$
\begin{gather}\label{rho}
\frac{1}{\rho^n}=\frac{1}{(n-1)!} \int_0^{\infty}  dt\, t^{n-1}
e^{-t\rho}, \qquad n>0,
\end{gather}
which follows from the Euler integral for the gamma-function. This
representation can be used to def\/ine the inverse d'Alembertian
\begin{gather}\label{boxinv}
\frac{1}{\Box}\,\psi=\int_0^{\infty} dt \,e^{-t\Box}\psi
\end{gather}
if a (vector-indices) function $\psi$ will be regarded as the sum
over the complete set of the d'Alembertian eigenfunctions $\psi_n$
with coef\/f\/icients $\alpha_n$
\begin{gather*}
\psi=\sum_n \alpha_n \psi_n.
\end{gather*}
By def\/inition
\begin{gather}\label{eigf}
\Box \psi_n=\lambda_n \psi_n
\end{gather}
with eigenvalues $\lambda_n$. Then, (\ref{boxinv}) and
(\ref{eigf}) result in
\begin{gather*}
\Box^{-1} \psi_n=\lambda^{-1}_n \psi_n,
\end{gather*}
and we have to complete the def\/inition of the inverse operator
to include the d'Alembertian zero modes. It could for instance be
done by replacing the eigenvalues with $\lambda^{\tt reg.}
_n=\lambda_n+\mathfrak{e}(\lambda,n)$ such that
$\lambda_n+\mathfrak{e}(\lambda,n)\ne 0$ \cite{gfs91}, and the
regularized d'Alembertian has a well def\/ined inverse operator.

To f\/igure out other features of the method, let us apply the
procedure of resolving to a~well-known case borrowed for instance
from supergravities. A simplest example is the Bianchi identity
for ${\rm D}=11$ supergravity gauge f\/ield \cite{bbs98,cjlp98II}
in the zero fermion setting
\begin{gather}\label{BSG}
dF^{(7)}=F^{(4)}\wedge F^{(4)};\qquad F^{(4)}=dA^{(3)},\qquad \ast
F^{(4)}=F^{(7)}.
\end{gather}
The r.h.s.\ of (\ref{BSG}) is a current form in our jargon, and it
is easy to see that the pre-current form is $G^{(7)}=A^{(3)}\wedge
F^{(4)}$. Hence the solution to the Bianchi identity is
$F^{(7)}=dA^{(6)}+A^{(3)}\wedge F^{(4)}$. Let us now recover this
result by use of (\ref{sol1}). It leads to
\begin{gather*}
F^{(7)}=d\chi^{(6)}+\Box^{-1}{\boldsymbol \delta}
\big(F^{(4)}\wedge F^{(4)}\big).
\end{gather*}
Noticing that $F^{(4)}\wedge F^{(4)}=d(A^{(3)}\wedge F^{(4)})$ we
have
\begin{gather*}
\Box^{-1}{\boldsymbol \delta} d \big(A^{(3)}\wedge
F^{(4)}\big)=\Box^{-1}(\Box-d{\boldsymbol \delta}) A^{(3)}\wedge
F^{(4)}= A^{(3)}\wedge F^{(4)}-d \Box^{-1}{\boldsymbol \delta}
\big(A^{(3)}\wedge F^{(4)}\big).\!\!
\end{gather*}
To get the last expression we have used $[\Box^{-1},d]=0$, which
comes from (\ref{dtri}) as
\[ [d,\Box]=0 \quad \leadsto \quad d=\Box d
\Box^{-1} \quad \leadsto \quad \Box^{-1}d=d \Box^{-1}.
\]
Finally,
\begin{gather*}
F^{(7)}=d \left( \chi^{(6)}-\Box^{-1}{\boldsymbol \delta}
\big(A^{(3)}\wedge F^{(4)}\big)\right)+A^{(3)}\wedge F^{(4)}\equiv
dA^{(6)}+A^{(3)}\wedge F^{(4)}.
\end{gather*}
The f\/inal result is a local expression since any trace of
non-locality is hidden with introducing a new f\/ield. As a
consequence, an appropriate local gauge transformation has to be
assigned to~$A^{(6)}$ rather than to $\chi^{(6)}$.

We would like to point out that one may encounter an opposite
situation when an initially local theory possesses an invariance
under special non-local transformations. An example of such
transformations is duality rotations in four-dimensional Maxwell
theory \cite{dt76}. A good choice of new variables may help to
avoid non-locality as we will see under discussing dif\/ferent
approaches to duality-symmetric theories.

\subsection{Dual to graviton f\/ield and its equation of motion}

At this stage we have all necessary tools to accomplish the
current task: equation (\ref{Rem4}) is resolved with
\begin{gather}\label{Fam}
\hat{F}^{[D-2]}_{\hat a}=d\hat{A}^{[D-3]}_{\hat
a}+\hat{\ast}\hat{\tilde{G}}^{[2]}_{\hat a}\equiv \hat{\mathbb
F}^{[D-2]}_{\hat a}+\hat{\ast}\hat{\tilde{G}}^{[2]}_{\hat a}
\end{gather}
(cf. equation (\ref{sol1})), where
\begin{gather*}
\hat{\ast}\hat{\tilde{G}}^{[2]}_{\hat a}=\hat{\square}
^{-1}\hat{\boldsymbol \delta}
\hat{\ast}\hat{\tilde{J}}^{[1]}_{\hat a}.
\end{gather*}
Here we have introduced the inverse to the
$\mathrm{D}$-dimensional d'Alembertian $\hat{\square}$ operator
together with the $\mathrm{D}$-dimensional co-derivative
$\hat{\boldsymbol \delta}$. Though the new gauge f\/ield
$\hat{A}^{[D-3]}_{\hat a}$ is the Hodge dual to the vielbein we
will nevertheless call them ``the graviton dual f\/ield''.

Equation (\ref{Fam}) is apparently non-local, but what is more
important for us is the locality of the dual f\/ield equation of
motion. This equation is
\begin{gather*}
d\left(\hat{\ast} d\hat{A}^{[D-3]}_{\hat a}\right)+d\hat{\tilde
G}^{[2]}_{\hat a}=0,
\end{gather*}
and since
\begin{gather*}
d \hat{\tilde G}^{[2]}_{\hat a}=d \hat{\ast}\left({\hat
\square}^{-1} \hat{\boldsymbol{\delta}} \hat{\ast}\hat{\tilde
J}^{[1]}_{\hat a}\right)=d\hat{\ast}\left({\hat \square}^{-1}
\hat{\boldsymbol \delta} d \hat{\tilde S}^{[D-2]}_{\hat a}\right)=
d\hat{\ast} \left( {\hat \square}^{-1} ({\hat \square}-d
\hat{\boldsymbol \delta})
\hat{S}^{[D-2]}_{\hat a} \right) \notag \\
\phantom{d \hat{\tilde G}^{[2]}_{\hat
a}}{}=d\hat{\ast}\hat{S}^{[D-2]}_{\hat a}-d\hat{\ast} {\hat
\square}^{-1} d \hat{\boldsymbol \delta} \hat{S}^{[D-2]}_{\hat
a}=d\hat{\ast}\hat{S}^{[D-2]}_{\hat a}-d\hat{\ast}\left(
d\left[{\hat \square}^{-1}\hat{\boldsymbol
\delta}\hat{S}^{[D-2]}_{\hat a}\right] \right) ,
\end{gather*}
we arrive at
\begin{gather*}
d \left( \hat{\ast} d\left(\hat{A}^{[D-3]}_{\hat a}-{\hat
\square}^{-1}\hat{\boldsymbol \delta}S^{[D-2]}_{\hat a}\right)
\right)=-d\hat{\ast}\hat{S}^{[D-2]}_{\hat a}.
\end{gather*}
Introducing a new f\/ield $\hat{\mathbb A}^{[D-3]}_{\hat
a}=\hat{A}^{[D-3]}_{\hat a}-{\hat \square}^{-1}\hat{\boldsymbol
\delta}S^{[D-2]}_{\hat a}$ we f\/inally get the local equation of
motion for the dual f\/ield
\begin{gather}\label{Abbem}
d\left(\hat{\ast} d\hat{\mathbb A}^{[D-3]}_{\hat
a}+\hat{\ast}\hat{S}^{[D-2]}_{\hat a}\right)=0.
\end{gather}

The next point we have to check is the number of physical degrees
of freedom which is described by the dual f\/ield. Since this
f\/ield possesses a gauge invariance under
\begin{gather*}
\delta \hat{\mathbb A}^{[D-3]}_{\hat
a}=d\hat{\alpha}^{[D-4]}_{\hat a},
\end{gather*}
f\/ixing the gauge and the on-shell local Lorentz invariance (see
a more detailed discussion below) results in
\begin{gather}\label{Adof}
\underbrace{D \times
(D-2)}_{\text{gauge~f\/ixing~on-shell}}-\underbrace{\frac{1}{2}D
\times (D-1)}_{\mathrm{local~Lorentz}}=\frac{1}{2}D \times (D-3)
\quad \mathrm{d.o.f.},
\end{gather}
which matches precisely the on-shell degrees of freedom of the
vielbein $\hat{e}^{\hat a}_{\hat m}$.

The equation of motion for the dual f\/ield has an odd structure
from the point of view of dynamics of self-interacting f\/ields.
But this peculiarity comes directly from the absence of the true
type dual f\/ield in our construction. Roughly speaking, if we
were deal with the true type graviton dual f\/ield it would be
realized in having self-interacting terms of the graviton dual
f\/ield in its equation of motion. Moreover, it would also be
natural to use another (generalized) Hodge star operator in the
equation like (\ref{Abbem}) that would correspond to a new type
metric tensor (or its generalization) constructed out of the true
type dual f\/ield. In the approach we follow, we deal instead with
a non-complete dualization when the self-interacting part of the
``dual'' f\/ield dynamics is replaced with a self-interaction of
vielbeins, and the Hodge star operator is def\/ined in the
standard manner. However, it does not mean that equation
(\ref{Abbem}) is a free f\/ield equation for the graviton dual:
the Hodge star clearly indicates that $\hat{\mathbb
A}^{[D-3]}_{\hat a}$ interacts with gravity.

\begin{remark}
The situation we encountered here is another side of numerous
no-go theorems on constructing the self-interacting dual models of
Yang--Mills theories and General Relativity
\cite{dt76,bbh03,bb03,bch03,abp05,ds05}. Drawbacks on this way
could be recognized just looking at the system of a Yang--Mills
equation of motion and the Bianchi identity
\begin{gather*}
D_{A}\ast F=0,\qquad
D_{A} F=0 .
\end{gather*}

Naively, it seems to be a symmetry that would be a generalization
of Abelian duality rotations that change $F$ with $\ast F$ and the
equation of motion with the Bianchi identity. In fact, $F
\leftrightarrow \tilde{F}\equiv \ast F$ results in
\begin{gather}
D_{A}\tilde{F}=0,\qquad D_{A} \ast{\tilde F}=0.\label{YMtr}
\end{gather}

But (\ref{YMtr}) is not the set of equations of the true
self-interacting dual system
\begin{gather*}
D_{\tilde A}\tilde{F}=0,\qquad
D_{\tilde A} \ast{\tilde F}=0,
\end{gather*}
where $D_{\tilde A}$ has to be the covariant with respect to the
true dual gauge f\/ield $\tilde{A}$ derivative, and the f\/ield
strength of $\tilde{A}$ is $\tilde{F}$.

This example shows that the def\/inition of duality (self-duality)
\begin{gather*}
\tilde{F}_{mn}=\frac{1}{2} \epsilon_{mnpq} F^{pq} \qquad
(\tilde{F}=F)
\end{gather*}
borrowed from the Maxwell theory does not guarantee the pass to
the dual (self-dual) description and even less meaningful for a
Yang--Mills theory.

All these facts are tightly related to a classical result by Wu
and Yang \cite{wy75} which consists of two statements: the f\/ield
strength of a non-Abelian f\/ield does not determine the gauge
f\/ield even locally in any small region, and that the dual of a
sourceless non-Abelian gauge f\/ield is not always also a gauge
f\/ield (see also \cite{dw76,dd79}). Another illustration of this
result is absence of duality rotations in a~Yang--Mills theory
\cite{dt76}.
\end{remark}

\section{Duality-symmetric action for General Relativity}

After having established the on-shell description of the dual
f\/ield, the next task is to f\/igure out an action from which
this dynamics follows. As we already discussed, the dual f\/ield
does interact with gravity. To produce this interaction we shall
take into account the dynamics of gravity. Then the action we seek
should contain at least the Einstein--Hilbert--Palatini Lagrangian
together with terms which manage the dynamics of the dual f\/ield
$\hat{\mathbb A}^{[D-3]}_{\hat a}$. Therefore, we have to deal
with a duality-symmetric formulation which is democratic with
respect to the f\/ields. In presence of dual partners the initial
degrees of freedom double, so to reduce it twice we have to impose
the duality relation
\begin{align}\label{dr}
&\hat{\ast}d\hat{e}_{\hat a}=\hat{F}^{[D-2]}_{\hat a}\equiv
\hat{\mathbb F}^{[D-2]}_{\hat a}+\hat{\ast}\hat{\tilde
G}^{[2]}_{\hat a}
\end{align}
at least by hands. Recall that in (\ref{dr})
\begin{gather*}
\hat{\mathbb F}^{[D-2]}_{\hat a}=d\hat{\mathbb A}^{[D-3]}_{\hat
a},\qquad \hat{\ast}\hat{\tilde G}^{[2]}_{\hat
a}=\hat{S}^{[D-2]}_{\hat a}+(-)^{\frac{D(D-5)}{2}}{\hat
\square}^{-1}\hat{\boldsymbol \delta}\hat{\ast}
\hat{J}^{[1]}_{\hat a},
\end{gather*}
the forms $\hat{S}^{[D-2]}_{\hat a}$ and $\hat{J}^{[1]}_{\hat a}$
have been introduced in (\ref{J}), (\ref{S}), and
\begin{gather*}
\hat{\mathbb{A}}^{[D-3]}_{\hat a}=\hat{A}^{[D-3]}_{\hat a}-{\hat
\square}^{-1}\hat{\boldsymbol \delta}S^{[D-2]}_{\hat a}.
\end{gather*}

It is not so straightforward to realize the explicit form of the
Lagrangian for the dual f\/ield. This is mainly due to the
vielbeins' interacting part $\hat{S}^{[D-2]}_{\hat a}$ in
(\ref{Abbem}). In principle, it is enough to get a formulation
from which (\ref{dr}) will follow as an equation of motion and as
a supplement to the Einstein equation. But it is still difficult
to deduce this additional part to the Einstein--Hilbert--Palatini
action from the very beginning. We propose to reach the goal
indirectly by use of the following road map:
\begin{gather*}
\xymatrix{ \hat{e}^{\hat a}_{\hat m} \ar[d]_{\text{KK reduction}}
&&&& \big(\hat{e}^{\hat a}_{\hat m}, {\hat{A}_{{\hat m}_1 \dots
{\hat m}_{D-3}}}^{\hat a}\big)
\\
\big(e^a_m,\phi,A^{[1]}\big)
\ar[rrrr]_{\phi,~A^{[1]}~\text{dualization}}&&&&
\big(\overbrace{e^a_m,\phi,A^{[1]}},\overbrace{A^{[D-4]},A^{[D-3]}}\big)
\ar[u]_{\text{KK oxidation}} }
\end{gather*}

 Drawing this picture we have taken into account
that a part of the dual to graviton f\/ield will become dual
f\/ields to a dilaton and to a Kaluza--Klein vector after
reduction of gravity from $\mathrm{D}$ to $\mathrm{D-1}$
dimensional theory. We are going to use this fact to deduce a
reduction ansatz for the graviton dual f\/ield strength which will
produce the correct structure of the duality-symmetric gravity in
$\mathrm{D-1}$ dimensions. After establishing the ansatz we will
apply it for the uplifting the known $\mathrm{D-1}$ dimensional
action to get the action in $\mathrm{D}$ dimensions.

\subsection{Dimensional reduction ansatz}

To pass through the abovementioned route we shall reduce
$\mathrm{D}$-dimensional gravity action f\/irst. Let us split the
indices as $(a,z)$ with $a=0,\ldots,D-2$ and choose the standard
Kaluza--Klein ansatz for vielbeins
\begin{gather}\label{KKe}
\hat{e}^{a[1]}=e^{\alpha \phi} e^{a[1]},\qquad
\hat{e}^{z[1]}=e^{\beta\phi}\big(dz+A^{[1]}\big),
\end{gather}
where $\phi$ denotes the dilaton f\/ield, $A^{[1]}$ stands for the
Kaluza--Klein vector f\/ield and $z$ is the direction of the
reduction. All the f\/ields on the r.h.s.\ of equation \eqref{KKe}
are independent on the reduction coordinate. The parameters
$\alpha$ and $\beta$ are related to the number of the space-time
dimensions and to each other as \cite{lpss95}
\begin{gather*}
\alpha^2=\frac{1}{2(D-2)(D-3)},\qquad \beta=-(D-3)\alpha,
\end{gather*}
that corresponds to the Einstein frame after reduction.

Using the standard technique of dimensional reduction
\cite{lpss95,popelec} one can verify that a set of equations that
come from the reduction of $\mathrm{D}$-dimensional Einstein
equation (\ref{Rem}) in a $\mathrm{D-1}$-dimensional zero-torsion
setting follows from the action
\begin{gather*}
S=\int_{{\mathcal M}^{D-1}} \left[ -R^{ab} \wedge
\Sigma_{ab}+\frac{1}{2} d\phi \wedge \ast d\phi -\frac{1}{2}
e^{-2(D-2)\alpha\phi}F^{[2]}\wedge \ast F^{[2]} \right],
\end{gather*}
where quantities without hats refer to $\mathrm{D-1}$ dimensions
and $F^{[2]}=dA^{[1]}$. These equations are
\begin{alignat}{5}\label{phiem}
&\frac{\delta {\mathcal L}}{\delta \phi}=0 \quad && \leadsto\quad
&& d(\ast d\phi)+(D-2)\alpha~ F^{[2]}\wedge
e^{-2(D-2)\alpha\phi}\ast
F^{[2]}=0,&\\
\label{Aem-1}
& \frac{\delta {\mathcal L}}{\delta A^{[1]}}=0 && \leadsto &&  d\big(e^{-2(D-2)\alpha\phi} \ast F^{[2]}\big)=0,&\\
\label{omem} & \frac{\delta {\mathcal L}}{\delta
\omega^{ab}}=0\quad  && \leadsto
&& \Sigma_{abc}\wedge T^c=0,&\\
\label{eom} & \frac{\delta {\mathcal L}}{\delta e^a}=0 && \leadsto
&& \Sigma_{abc} \wedge R^{bc}-(-)^{D-4} M^{[D-2]}_a=0.&
\end{alignat}
The latter equation involves the energy-momentum tensor of the
dilaton and of the Kaluza--Klein vector f\/ield which is def\/ined
by
\begin{gather*}
M^{[D-2]}_a\equiv \frac{\delta}{\delta e^a} (-)^{D-4} \left(
\frac{1}{2} d\phi \wedge \ast d\phi - \frac{1}{2}
e^{-2(D-2)\alpha\phi} F^{[2]} \wedge \ast F^{[2]} \right).
\end{gather*}

It is easy to see that the second order equations (\ref{phiem})
and (\ref{Aem-1}) can be recast in the f\/irst order Bianchi
identities for (Hodge) dual to $\phi$ and $A^{[1]}$ f\/ields,
which are def\/ined by the following duality relations
\begin{gather}\label{drphi}
\ast d\phi=-dA^{[D-3]}+(D-2)\alpha~ dA^{[D-4]}\wedge A^{[1]},
\\
\label{drA1} e^{2(\beta-\alpha)\phi} \ast dA^{[1]}=-dA^{[D-4]}.
\end{gather}

To reach a $\mathrm{D-1}$-dimensional analog of (\ref{Rem3}), one
has to resolve the torsion free condition~(\ref{omem}) and to
substitute the solution into the Einstein equation (\ref{eom}). In
ef\/fect, we get the following equation of motion for vielbeins
\begin{gather}
d(\ast d e_a)=-\ast{\mathfrak J}^{[1]}_a,\nonumber
\\
\label{frJ} {\mathfrak J}^{[1]}_a=(-)^{\frac{(D-1)(D-6)}{2}}
J^{[1]}_a-\ast dS^{[D-3]}_a,
\end{gather}
and
\begin{gather*}
J^{[1]}_a=\ast \left(\omega^{bc}(e) \wedge d
\Sigma_{abc}+(-)^{D-4} {\omega^b}_{d}(e) \wedge \omega^{dc}
(e)\wedge \Sigma_{abc}+M^{[D-2]}_a \right),\nonumber
\\
S^{[D-3]}_a=\ast (e^b \wedge e_a)~e^m_c e^n_b \partial_{[m}
e^c_{n]}.
\end{gather*}

Hence a $\mathrm{D-1}$-dimensional analog of (\ref{dr}) becomes
\begin{gather}\label{drvil}
\ast de_a=-(dA^{[D-4]}_a+\ast {\mathfrak G}^{[2]}_a),
\end{gather}
where
\begin{gather*}
\ast \mathfrak{G}^{[2]}_a=\Box^{-1} \boldsymbol{\delta} \ast
\mathfrak{J}^{[1]}_a.
\end{gather*}

Following the same calculations resulted in (\ref{Abbem}) one gets
the equation of motion of the dual f\/ield $A^{[D-4]}_a$
\begin{gather*}
d(\ast d\mathbb{A}^{[D-4]}_a-\ast S^{[D-3]}_a)=0,
\end{gather*}
where $\mathbb{A}^{[D-4]}_a$ is the redef\/ined f\/ield
\begin{gather*}
\mathbb{A}^{[D-4]}_a=A^{[D-4]}_a+\Box^{-1} \boldsymbol{\delta}
S^{[D-3]}_a.
\end{gather*}
In terms of the latter the duality relation (\ref{drvil}) becomes
\begin{gather}\label{drvil1}
\ast de_a=-\left(d \mathbb{A}^{[D-4]}_a +\ast
\mathfrak{G}^{[2]}_a\right),\qquad \ast
\mathfrak{G}^{[2]}_a=(-)^{\frac{(D-1)(D-6)}{2}}
\Box^{-1}\boldsymbol{\delta} J^{[1]}_a -S^{[D-3]}_a.
\end{gather}

So, we have derived the duality relations between
$\mathrm{D-1}$-dimensional f\/ields. Let us now turn to the
question what an ansatz for the $\mathrm{D}$-dimensional graviton
dual f\/ield should be to recover relations (\ref{drphi}),
(\ref{drA1}) and (\ref{drvil1}) directly from (\ref{dr})?

By construction the second term entering the r.h.s.\ of (\ref{dr})
is a function of vielbeins, hence its reduction is completely
determined by the standard Kaluza--Klein ansatz. The same is also
true for the l.h.s.\ of (\ref{dr}). To reduce the f\/irst term on
the r.h.s.\ of (\ref{dr}) we propose the following ansatz
(see~\cite{ajn04} for details)
\begin{gather}
\hat{\mathbb F}^{a [D-2]}= e^{2(D-3)\alpha\phi}\ast g^{a
[1]}\nonumber \\
\phantom{\hat{\mathbb F}^{a
[D-2]}=}{}+e^{-\alpha\phi}\left(d\mathbb{A}^{a
[D-4]}+\ast{\mathfrak G}^{a [2]}+\alpha \ast (d\phi\wedge
e^a)+e^{-\alpha\phi}\ast G^{a [2]}\right)\wedge
\big(dz+A^{[1]}\big),\label{Fa9s}
\\
\hat{\mathbb F}^{z
[D-2]}=-e^{-\beta\phi}\left(dA^{[D-3]}-(D-2)\alpha~
dA^{[D-4]}\wedge A^{[1]}\right)-e^{2(D-3)\alpha\phi+\beta\phi}
(1-\beta)\ast
d\phi\nonumber\\
\phantom{\hat{\mathbb F}^{z[D-2]}=}{}+e^{2(D-3)\alpha\phi}\ast
g^{z[1]}+ e^{-\beta
\phi}\left(dA^{[D-4]}+e^{-2\alpha\phi+\beta\phi}\ast
G^{z[2]}\right)\wedge \big(dz+A^{[1]}\big).\label{Fz9s}
\end{gather}
Here we have used the fact that under reduction
\begin{gather*}
\hat{\tilde G}^{[2]}_a=G^{[2]}_a+g^{[1]}_a\wedge
\big(dz+A^{[1]}\big),
\\
\hat{\tilde G}^{[2]}_z=G^{[2]}_z+g^{[1]}_z \wedge
\big(dz+A^{[1]}\big),
\end{gather*}
and $A^{[D-3]}$, $A^{[D-4]}$ f\/ields come from $\hat{A}^{z
[D-3]}$ component of the f\/ield $\hat{\mathbb A}^{{\hat a}
[D-3]}$. It is easy to check that the ansatz does lead to the
correct duality relation after the reduction of (\ref{dr}).

As we have mentioned above, we are going to use the reduction
ansatz to construct the action from which duality relations will
follow as equations of motion. To handle this problem let us
discuss f\/irst dif\/ferent ways to its possible solution.

\subsection{Non-covariant and covariant approaches to duality-symmetric theories}

There are two dif\/ferent ways of describing classical gauge
f\/ields. One of them is to describe a gauge theory in terms of
appropriate f\/ields, which are naturally associated with a theory
(antisymmetric tensor gauge f\/ields for a gauge system, spin-2
f\/ield for gravity etc.), another way is to make it in terms of
their dual partners, which possess the same number of dynamical
degrees of freedom on-shell.

To construct a formulation where f\/ields and their dual partners
would be on an equal footing, but without spoiling the initial
degrees of freedom, one should double the f\/ields with their
duals and to imply duality relations which have to decrease twice
the degrees of freedom of a doubled f\/ields system. Such a
formulation is called duality-symmetric. In some cases when a
gauge f\/ield satisf\/ies the self-duality condition, the
duality-symmetric formulation describes solely a single chiral (or
self-dual) f\/ield which has twice less dynamical degrees of
freedom in compare with its non-chiral companion.

Approaches which are used to describe self-dual f\/ields could be
generically divided into two categories: ones which sacrif\/ice
covariance in favor of manifest duality, and manifestly covariant
approaches. We will brief\/ly review both of them.

As an example of manifestly dual but non-covariant approach let us
consider the duality-symmetric formulation of $\mathrm{D}=4$
Maxwell theory \cite{dt76}.

Equations of ${\rm D}=4$ electrodynamics without sources
\begin{alignat*}{3}
& \frac{\partial \boldsymbol{E}}{\partial t}=\triangledown \times
\boldsymbol{B},\qquad &&  \triangledown \cdot \boldsymbol{E}=0, &\nonumber \\
& \frac{\partial \boldsymbol{B}}{\partial t}=-\triangledown \times
\boldsymbol{E},\qquad &&  \triangledown \cdot \boldsymbol{B}=0,& 
\end{alignat*}
are invariant under duality rotations
\begin{gather}\label{drot}
\delta {\boldsymbol E}=\alpha \boldsymbol{B}, \qquad \delta
{\boldsymbol B}=-\alpha \boldsymbol{E},
\end{gather}
while the Lagrangian
\begin{gather*}
\mathcal{L}=\frac{1}{2} \left[ \boldsymbol{E}^2-\boldsymbol{B}^2
\right]
\end{gather*}
is apparently not. Since $\boldsymbol{B}=\triangledown \times
\boldsymbol{A}$ is the solution to the Bianchi identity
$\triangledown \cdot \boldsymbol{B}=0$, rotations~(\ref{drot})
become spatially non-local in terms of $\boldsymbol{E}$ and
$\boldsymbol{A}$
\begin{gather*}
\delta {\boldsymbol E}=\alpha~ \triangledown \times
\boldsymbol{A}, \qquad \delta {\boldsymbol A}=\alpha~
\triangle^{-1} ( \triangledown \times \boldsymbol{E}).
\end{gather*}
Here $\triangle^{-1}$ is the inverse to the three-dimensional
Laplacian operator.

Turning to the Hamiltonian, one can see that
\begin{gather}\label{H}
\mathcal{H}=\frac{1}{2} \left( \boldsymbol{E}^2+\boldsymbol{B}^2
\right)-A_0~ \triangledown \cdot \boldsymbol{E},
\end{gather}
and, from the point of view of the Hamiltonian approach,
$\triangledown \cdot \boldsymbol{E}=0$ is a constraint which
enters~(\ref{H}) via the Lagrange multiplier $A_0$. One can
resolve this constraint as $\boldsymbol{E}=\triangledown \times
\boldsymbol{Z}$ and substitute this solution to the Hamiltonian
back. After that the Hamiltonian transforms into
\begin{gather*}
\mathcal{H}=\frac{1}{2}\left( (\triangledown \times
\boldsymbol{Z})^2+(\triangledown \times \boldsymbol{A})^2 \right).
\end{gather*}

{\samepage In terms of the new variables $\boldsymbol{A}$ and
$\boldsymbol{Z}$ duality rotations are local
\begin{gather}\label{drotl}
\delta {\boldsymbol Z}=\alpha~\boldsymbol{A}, \qquad \delta
{\boldsymbol A}=-\alpha~ \boldsymbol{Z},
\end{gather}
and the Hamiltonian is apparently invariant under (\ref{drotl}).
But what about the Lagrangian?}

To tackle this question, it is convenient to consider the
Lagrangian in the f\/irst order form
$\mathcal{L}=p\dot{q}-\mathcal{H}$. Then the Lagrangian becomes
\begin{gather*}
\mathcal{L}=\dot{\boldsymbol{A}} \triangledown \times
\boldsymbol{Z}- \frac{1}{2}\left( (\triangledown \times
\boldsymbol{Z})^2+(\triangledown \times \boldsymbol{A})^2 \right),
\end{gather*}
or
\begin{gather*}
\mathcal{L}=\frac{1}{2}\epsilon_{ab}\dot{\boldsymbol{Z}^b}
(\triangledown \times \boldsymbol{Z}^a)- \frac{1}{2}\delta_{ab}
(\triangledown \times \boldsymbol{Z}^a) (\triangledown \times
\boldsymbol{Z}^b) \nonumber\\ \phantom{\mathcal{L}=}{}
+(\mathrm{total~time~derivative})+(\mathrm{total~spatial~derivative}),
\end{gather*}
where we have joined two potentials into
$\boldsymbol{Z}^a=(\boldsymbol{Z},\boldsymbol{A})$. If we will
chose boundary conditions to be good enough to get rid of total
derivatives, then the Lagrangian will also be invariant under
duality rotations (\ref{drotl}). This form of the non-covariant
f\/irst-order Lagrangian proposed in \cite{dt76} was re-derived
later in \cite{ss94} within the pure Lagrangian framework.

\begin{remark}
The f\/irst duality-symmetric formulation of ${\rm D}=4$ Maxwell
theory with two gauge potentials was proposed in \cite{z71}. A
benef\/it of the approaches \cite{z71} and \cite{dt76} is  having
the manifest duality symmetry. However, to reach the formulation
of \cite{dt76} constraints have to be resolved. It is not a
problem for an Abelian theory, but it becomes a problem for a
non-Abelian gauge theory as well as for General Relativity.

It is also noteworthy to point out that the formulation of
\cite{ss94} is f\/lexible enough to describe not only self-dual
f\/ields which possess an invariance under duality rotations, but
also duality-symmetric f\/ields, dual to each other in the Hodge
sense, and do not possessing an invariance under duality
transformations like (\ref{drot}). The formalism of \cite{dt76} is
mostly well-suited to cases with self-dual f\/ields. Another
advantage of \cite{ss94} is a possibility to covariantize the
description as we will see in a minute.
\end{remark}

Progress towards the Lagrangian covariant description of self-dual
gauge f\/ields was hampered by the no-go theorem of~\cite{ms82}.
This caused problems with consistent coupling of self-dual
f\/ields to gravity and other f\/ields, which was important in the
context of constructing the type IIB supergravity action with a
rank four self-dual f\/ield, as well as for searching for
ef\/fective actions for super-p-branes with self-dual f\/ields on
their worldvolume in a supergravity background. However, in the
end, there were proposed covariant actions for duality-symmetric
theories which overcame the no-go theorem by drawing auxiliary
f\/ields into the construction.

Let us demonstrate a way of keeping covariance with the following
example of the duality-symmetric Maxwell theory in $\mathrm{D}=3$
space-time dimensions. The Maxwell f\/ield action
\begin{gather*}
S=\int  d^3 x\, \frac{1}{4} F_{mn} F^{mn}, \qquad F_{mn}=2
\partial_{[m} A_{n]}
\end{gather*}
determines the gauge f\/ield dynamical equation
\begin{gather}\label{Aeom3}
\partial_m F^{mn}=0,
\end{gather}
and due to the gauge invariance, $A_m$ contains one degree of
freedom. The same number of degrees of freedom can be associated
with a scalar f\/ield $\varphi$, which is dual to the original
gauge f\/ield. The duality between f\/ields means, in particular,
that one can present the equation of motion~(\ref{Aeom3}) as the
Bianchi identity for the dual f\/ield, and this Bianchi identity
(as well as the original gauge f\/ield equation of motion) follows
from the duality relation
\begin{gather}\label{dr3}
F_{mn}-\epsilon_{mnk} \partial^k \varphi=0.
\end{gather}
Indeed, applying $\partial^m$ to (\ref{dr3}) leads to the equation
of motion (\ref{Aeom3}) since
\begin{gather*}
\partial^m \epsilon_{mnk} \partial^k \varphi=0
\end{gather*}
is the Bianchi identity for the scalar f\/ield.

However, the duality relation (\ref{dr3}) contains also the
dynamics of the dual f\/ield. One can present (\ref{dr3}) in the
equivalent form
\begin{gather*}
\epsilon^{mnk}F_{mn}-2\partial^k \varphi=0,
\end{gather*}
and acting on the latter with $\partial_k$ leads now to the
equation of motion for the dual f\/ield
\begin{gather*}
\partial_m \partial^m \varphi=0
\end{gather*}
since
\begin{gather*}
\epsilon^{mnk} \partial_m F_{nk}=0
\end{gather*}
is the Bianchi identity for the original gauge f\/ield.

Hence, one can describe the original system in the
duality-symmetric manner doubling the f\/ields and thus having
$1+1=2$ degrees of freedom, that will reduce to one initial degree
of freedom after imposing the duality relation between dual
partners. Indeed, the dynamics of the doubled f\/ield system is
def\/ined by the following set of equations
\begin{gather}
F^{mn}-\epsilon^{mnk}\partial_k \varphi =0,\nonumber\\
\partial_m \partial^m \varphi=0,\nonumber\\
\partial_m F^{mn}=0,\label{eomtr3}
\end{gather}
and depending on our purposes we can choose between describing the
model in terms of the vector potential $A_m$, of the scalar
$\varphi$, or in terms of both f\/ields when we describe them in
the duality-symmetric manner.

It is easy to recover the action which generates the dynamics of
dual f\/ields in the case, however we would like to f\/ind an
action from which it could be possible to derive the duality
relation as well. A simplest way is to incorporate (\ref{dr3})
into the action with help of a Lagrange multiplier
\begin{gather}\label{dsM3}
S=\int  d^3 x \left[ \frac{1}{4} F_{mn} F^{mn}+\frac{1}{2}
\partial_m \varphi \partial^m \varphi +\Lambda_{(1) mn}
\big(F^{mn}-\epsilon^{mnk}\partial_k \varphi\big)\right ],
\end{gather}
but the Lagrange multiplier should not carry new degrees of
freedom. In practice we have an opposite situation: the Lagrange
multiplier propagates and there is no additional symmetry to kill
its dynamical degrees of freedom. To observe the Lagrange
multiplier propagation it is enough to take a look at equations of
motion that come from (\ref{dsM3})
\begin{gather}
F^{mn}-\epsilon^{mnk}\partial_k \varphi =0,\nonumber\\
\partial_m \partial^m \varphi + \partial_m \epsilon^{mnk}
\Lambda_{(1)nk}=0,\nonumber\\
\partial_m F^{mn}+\partial_m \Lambda_{(1)}^{mn}=0,\label{eom3}
\end{gather}
and to compare this set to that of the duality-symmetric theory
(\ref{eomtr3}). It is easy to see that~(\ref{eom3}) recasts in
(\ref{eomtr3}) if\/f
\begin{gather*}
\partial_m \Lambda_{(1)}^{mn}=0,\qquad
\partial_m \epsilon^{mnk}
\Lambda_{(1)nk}=0,
\end{gather*}
that means in its turn
\begin{gather*}
\Lambda_{(1)mn}=2\partial_{[m} B_{(1) n]}.
\end{gather*}
Hence $\Lambda_{(1)mn}$ corresponds to a propagating f\/ield.

We can try to correct such a situation with introducing a new
Lagrange multiplier, varying over which will lead to setting
$\Lambda_{(1)mn}$ to zero. That is consider the action
\begin{gather*}
S=\int  d^3 x \left[ \frac{1}{4} F_{mn} F^{mn}+\frac{1}{2}
\partial_m \varphi \partial^m \varphi +\Lambda_{(1) mn}
\big(F^{mn}-\epsilon^{mnk}\partial_k \varphi\big)\right.\nonumber\\
\left.\phantom{S=}{}-\frac{1}{2} \Lambda_{(1)
mn} \Lambda^{(1) mn}+\Lambda_{(2) mn}\Lambda^{(1) mn} \right].
\end{gather*}
Variation over $\Lambda_{(2) mn}$ results in $\Lambda_{(1) mn}=0$.
But the same analysis as before demonstrates that the new Lagrange
multiplier is dynamical and there is no additional symmetry to get
rid of the new degrees of freedom.

We can continue this process further on, and it never stops.
Hence, the action we will get in the end is
\begin{gather*}
S=\int  d^3 x  \left[ \frac{1}{4} F_{mn} F^{mn}+\frac{1}{2}
\partial_m \varphi \partial^m \varphi +\Lambda_{(1) mn}
\big(F^{mn}-\epsilon^{mnk}\partial_k \varphi\big)-\frac{1}{2}
\Lambda_{(1)
mn} \Lambda^{(1) mn}\right] \nonumber \\
\phantom{S=}{}+\int  d^3 x ~ \sum^{\infty}_{n=0} (-)^n
\Lambda_{(n+1)
mn}\Lambda^{(n+2) mn},
\end{gather*}
and this is a manifestly Lorentz covariant action constructed out
of an inf\/inite tail of Lagrange multipliers
\cite{myw90,w91,mr94,bk97,b97}.

However, there is a way \cite{pst95plb,pst95prd,pst97prd} to cut
the inf\/inite tail at an $N+1$-th step replacing the $N$-th
Lagrange multiplier with
$\partial_{[n}a(x)\Lambda_{(0)m]k}\partial^k a(x)$. Then, in the
simplest case of $N=1$, the action becomes
\begin{gather*}
S=\int  d^3 x  \left[ \frac{1}{4} F_{mn} F^{mn}+\frac{1}{2}
\partial_m \varphi \partial^m \varphi +\Lambda_{(1) mn}
\big(F^{mn}-\epsilon^{mnk}\partial_k \varphi\big)-\frac{1}{2}
\Lambda_{(1)
mn} \Lambda^{(1) mn}\right] \nonumber \\
\phantom{S=}{}+\int  d^3 x \Lambda_{(2)}^{mn}\left(\Lambda_{(1)
mn}-\Lambda_{(0)mp}\partial^p a \partial_n a \right).
\end{gather*}
One can solve algebraic equations of motion of Lagrange
multipliers to f\/inally get
\begin{gather}\label{pstac3}
S=\int  d^3 x  \left[ \frac{1}{4} F_{mn} F^{mn}+\frac{1}{2}
\partial_m \varphi \partial^m \varphi-\frac{1}{2(\partial
a)^2}\partial_m \mathcal{F}^{mn} \mathcal{F}_{np} \partial^p a
\right ].
\end{gather}
Here $\mathcal{F}^{mn}=F^{mn}-\epsilon^{mnk}\partial_k \varphi$,
and (\ref{pstac3}) is the action originally proposed in
\cite{pst95plb,pst95prd,pst97prd}. Note the appearance of the new
f\/ield $a(x)$ which ensures the covariance of the model.

Let us compare the equations of motion coming from the PST action
(\ref{pstac3}) with that of~(\ref{eomtr3}). Varying the action
over $A_m$ and $\varphi$ results in
\begin{gather}
\partial_m F^{mn}-\partial_m\left( 2\frac{1}{(\partial a)^2}
\partial^{[m}a \mathcal{F}^{n]p}\partial_p a \right)=0,\nonumber\\
\partial_m \partial^m \varphi+\epsilon^{mnk} \partial^k
\left(\frac{1}{(\partial a)^2}
\partial^{[m}a \mathcal{F}^{n]p}\partial_p a \right)=0.\label{PSTeom3}
\end{gather}
Clearly, if the generalized f\/ield strength $\mathcal{F}_{mn}$
were satisf\/ied $\mathcal{F}_{mn}=0$ on the mass shell, it would
produce the duality relation (\ref{dr3}), and the set of equations
of motion following from the PST action would be the same as
(\ref{eomtr3}). Moreover, it is enough to get solely the duality
relation from the action since the dynamics of f\/ields is encoded
within. Another way to reach the same conclusion is to take into
account the following identity
\begin{gather}\label{idD3}
\frac{1}{(\partial a)^2}
\partial^{[m}a \mathcal{F}^{n]p}\partial_p
a=\frac{1}{2}\mathcal{F}^{mn}-\frac{3}{2} \frac{1}{(\partial a)^2}
\partial^{[m}a \mathcal{F}^{np]}\partial_p a,
\end{gather}
and to write down (\ref{PSTeom3}) as
\begin{gather}\label{PSTD3eomm}
\partial_m \left( \frac{1}{(\partial a)^2}
\partial^{[m}a \mathcal{F}^{np]}\partial_p
a \right)=0,\qquad \epsilon_{mnk} \partial^k \left(
\frac{1}{(\partial a)^2}
\partial^{[m}a \mathcal{F}^{np]}\partial_p
a \right)=0.
\end{gather}
Obviously, $\mathcal{F}_{mn}=0$ is at least a particular solution
to (\ref{PSTD3eomm}), and we can use the duality relation as a
generating dynamical equation.

In fact, one can establish more: the latter system of equations
can be exactly reduced to the duality relation
$\mathcal{F}_{mn}=0$ \cite{pst95plb,pst95prd,pst97prd}. This is
provided by one of the two special symmetries of the action
(\ref{pstac3}), the so-called PST symmetries, and we will
demonstrate it for our model of the duality-symmetric gravity
below in details. The other PST symmetry serves for eliminating
one additional degree of freedom coming with the PST scalar
f\/ield $a(x)$. It breaks manifest covariance and establishes the
connection to the approach of \cite{ss94}. Dualization of the PST
scalar f\/ield \cite{msp99} relates the PST approach to that of
\cite{z71} and in such a way the latter formulation gets related
to \cite{dt76,ss94}. Other features of manifestly covariant
duality-symmetric models can be found in a recent review
\cite{s02}.

So far we outlined various methods of constructing the Lagrangians
for duality-symmetric theories, and we have to f\/igure out the
approach which would be more well-suited to the model we describe.
We would like the approach to be independent on space-time
dimensions and manifestly covariant. These requirements select the
method of \cite{pst95plb,pst95prd,pst97prd} as more economic in
compare with introducing the inf\/inite tail of Lagrange
multipliers, so we will follow this way.

\subsection{Duality-symmetric action and its symmetries}

After having discussed prerequisites let us focus on the
construction of the action. This procedure requires more tuning in
comparison to the on-shell description, so we are f\/ixing the
space-time dimensions $\mathrm{D}$ to be eleven. After dimensional
reduction to ten dimensions we should, in particular, reproduce
from the action the duality relations \eqref{drphi}, \eqref{drA1}
which now have the following form
\begin{gather}\label{F8dr}
{\mathcal F}^{[8]}=dA^{[7]}+e^{-\frac{3}{2}\phi}\ast dA^{[1]}=0,
\\
\label{F9dr} {\mathcal
F}^{[9]}=\left(dA^{[8]}-\frac{3}{4}dA^{[7]}\wedge
A^{[1]}\right)+\ast d\phi=0.
\end{gather}
Recall that (\ref{F8dr}) and (\ref{F9dr}) are the duality
relations between the scalar f\/ield $\phi$ and the Kaluza--Klein
vector f\/ield $A^{[1]}$ and their dual f\/ields $A^{[8]}$ and
$A^{[7]}$. In the above we have also f\/ixed the parameters
$\alpha$ and $\beta$ to be
\begin{gather*}
\alpha=+\frac{1}{12},\qquad \beta=-\frac{2}{3}.
\end{gather*}
The relevant part of the ten-dimensional duality-symmetric action
from which these relations come is (cf.\ \cite{bns})
\begin{gather}\label{dsL10}
S_{d.s.}=\frac{1}{2} \int_{{\mathcal M}^{10}}
\sum_{n=1}^{2}\left(F^{[10-n]}\wedge F^{[n]}+i_v {\mathcal
F}^{[10-n]}\wedge v \wedge F^{[n]}+v\wedge F^{[10-n]}\wedge i_v
{\mathcal F}^{[n]}\right).
\end{gather}

Our aim is to demonstrate that the action \eqref{dsL10} follows
after dimensional reduction from the part of the duality-symmetric
gravity action
\begin{gather}\label{dsgL11}
S_{\rm PST}=\int_{{\mathcal M}^{11}} \frac{1}{2} \hat{v}\wedge
\hat{\mathcal F}^{{\hat a}[9]}\wedge i_{\hat v} \hat{\mathcal
F}^{{\hat b}[2]}\eta_{{\hat a}{\hat b}},
\end{gather}
where we have introduced a scalar f\/ield $a(\hat{x})$
\cite{pst95plb,pst95prd,pst97prd} that enters the action in a
non-polynomial way through the one-form ${\hat v}$
\begin{gather*}
\hat{v}=\frac{d a(\hat{x})}{{\sqrt{-\partial_{\hat m}a~
\hat{g}^{{\hat m}{\hat n}}~\partial_{\hat n}a}}},
\end{gather*}
$i_{\hat v} \hat{\Omega}$ denotes the contraction of $\hat{v}$
with a dif\/ferential form $\hat{\Omega}$ (see Appendix A),
$\eta_{{\hat a}{\hat b}}$ is the Minkowski metric tensor and
\begin{gather}\label{F2a11}
\hat{\mathcal F}^{{\hat a}[2]}=d\hat{e}^{{\hat
a}}-\hat{\ast}\left(d\hat{\mathbb A}^{{\hat
a}[8]}+\hat{\ast}\hat{\tilde G}^{{\hat a}[2]}\right),
\\
\label{F9a11} \hat{\mathcal F}^{{\hat
a}[9]}=-\hat{\ast}\hat{\mathcal F}^{{\hat a}[2]}.
\end{gather}
Note that though these generalized f\/ield strengths are
constructed out of non-covariant quantities, we shall require them
to enter (\ref{F2a11}), (\ref{F9a11}) in the covariant
combinations.

\begin{remark}
Let us discuss the covariance of $\hat{\mathcal F}^{{\hat a}[2]}$
and $\hat{\mathcal F}^{{\hat a}[9]}$ in more detail. We are
dealing with objects in the dif\/ferential forms notation, hence
we do not worry about covariance under the general coordinate
transformations. However, we have a Lorentz index which has to be
rotated under the local Lorentz transformations. Vielbeins and a
spin connection are transformed as
\begin{gather*}
\hat{e}^{\prime}={\hat e} \hat{\Lambda} \quad \leftrightsquigarrow
\quad \hat{e}^{\prime {\hat a}}={\hat e}^{\hat b}
{\hat{\Lambda}_{\hat b}}^{\hat a},\nonumber \\
{\hat \omega}^{\prime}=\hat{\Lambda}^{-1}\hat{\omega}
\hat{\Lambda}+\hat{\Lambda}^{-1} d \hat{\Lambda} \quad
\leftrightsquigarrow \quad \hat{\omega}^{\prime
\hat{a}\hat{b}}={(\hat{\Lambda}^{-1})^{\hat a}}_{\hat c}
\hat{\omega}^{\hat{c}\hat{d}} {\hat{\Lambda}_{\hat d}}^{\hat
b}+{(\hat{\Lambda}^{-1})^{\hat a}}_{\hat c} d
\hat{\Lambda}^{\hat{c}\hat{b}},
\end{gather*}
that induces the following transformations of torsion and
curvature
\begin{gather*}
\hat{T}^{\prime}=\hat{T} \hat{\Lambda},\qquad
\hat{R}^{\prime}=\hat{\Lambda}^{-1} \hat{R} \hat{\Lambda}.
\end{gather*}
They are transformed as true tensors and they are covariant
objects.

On the other hand, equation (\ref{Rem3}) has to be covariant under
the local Lorentz transformations since it follows from the
Einstein equation, which is constructed out of true tensors.
However, the l.h.s.\ of (\ref{Rem3}) does not transform as a~true
tensor; it has an additional part depending on derivatives of
$\Lambda$. But as it was mentioned before, the r.h.s.\ of the same
equation is also not a~true tensor, and the additional
non-tensorial part of its transformation has to compensate the
undesirable terms on the l.h.s.

To be precise,
\begin{gather*}
\big[ d(\hat{\ast} d{\hat e})-\hat{\ast}\hat{\tilde J}^{[1]} \big
]^{\prime}= \big[ d(\hat{\ast} d{\hat e})-\hat{\ast}\hat{\tilde
J}^{[1]}\big ]\cdot \hat{\Lambda},
\end{gather*}
that means
\begin{gather}\label{dFL}
\big[ d \hat{\mathcal{F}}^{[9]} \big]^{\prime}=\big[ d \hat{
\mathcal{F}}^{[9]} \big] \cdot \hat{\Lambda}.
\end{gather}

To require the invariance of (\ref{dsgL11}) under the local
Lorentz transformations, we have to require the covariance of
$\hat{\mathcal{F}}^{[9]}$, i.e.
\begin{gather}\label{FL}
\big[ \hat{\mathcal{F}}^{[9]} \big]^{\prime}=\big[  \hat{
\mathcal{F}}^{[9]} \big] \cdot \hat{\Lambda}.
\end{gather}
At a f\/irst glance the latter does not seem possible because of
an apparent contradiction when we compare (\ref{FL}) with
(\ref{dFL}). But we still have a freedom in assigning a gauge
transformation to $\hat{\mathbb{A}}^{[8]}$. Taking such a
transformation in the form
\begin{gather}\label{AL}
\big[ d\hat{\mathbb A}^{[8]} \big]^{\prime}=d\big( \hat{\mathbb
A}^{[8]} \cdot \hat{\Lambda} \big)+d^{-1} \big[ \hat{\mathcal
F}^{[9]}\wedge d \hat{\Lambda} \big]
\end{gather}
will formally solve the problem. Here we have introduced the
inverse to $d$ operator whose action on an arbitrary form is
def\/ined via Schwingers' proper-time trick
\begin{gather}\label{d-1}
d^{-1} \omega^{[p]}=\int_0^{\infty}  d\tau \, e^{-\tau d}\,
\omega^{[p]},
\end{gather}
or, equivalently, through equation (\ref{rho}). Clearly, this
operator cannot be determined for exact forms, since
$dd^{-1}=d^{-1}d=\text{id}$ and acting on $d\Omega=0$ with
$d^{-1}$ results in $\Omega=0$, $\forall\; \Omega$ that is not
true of course. We have emphasized that the inverse to the
d'Alembertian operator is determined on a space of forms modulo
the d'Alembertian zero modes. The same is relevant for~(\ref{d-1}),
 since exact forms are zero modes of $d$. However, as
in the former case, one can regularize the external derivative
operator to include its zero modes into consideration. It also has
to be emphasized that one shall take care in treating $d^{-1}$,
taking $dd^{-1}=d^{-1}d=\text{id}$ f\/irst. For instance, the
expression $ddd^{-1}\cdots$ has to be equal to $d\cdots$ rather
than to zero.

Although the transformation (\ref{AL}) is highly non-local, it is
reduced on shell to the standard local Lorentz transformation what
we have used in (\ref{Adof}).
\end{remark}

To reduce the action \eqref{dsgL11}, we split of\/f the eleventh
coordinate $z$ from the ten-dimensional tangent space indices
$a=0,\dots,9$ after that we can rewrite the former action as
\begin{gather*}
S_{\rm PST}=\int_{{\mathcal M}^{11}} \frac{1}{2} \hat{v}\wedge
\hat{\mathcal F}^{a [9]}\wedge i_{\hat v} \hat{\mathcal F}^{ b
[2]}\eta_{ab}-\frac{1}{2}\hat{v}\wedge \hat{\mathcal F}^{z
[9]}\wedge i_{\hat v} \hat{\mathcal F}^{ z [2]}.
\end{gather*}
By use of ans\"atze \eqref{Fa9s}, \eqref{Fz9s} one gets
\begin{gather*}
\hat{\mathcal F}^{a [2]}=e^{\frac{1}{12}\phi}\big[de^a+\ast
\big(d{\mathbb A}^{a [7]}+\ast \mathfrak{G}^{a
[2]}\big)\big]\equiv e^{\frac{1}{12}\phi}{\mathcal F}^{a
[2]},\nonumber
\\
\hat{\mathcal F}^{a [9]}=e^{-\frac{1}{12}\phi}{\mathcal F}^{a
[8]}\wedge \big(dz+A^{[1]}\big),\qquad {\mathcal F}^{a [8]}=\ast
{\mathcal F}^{a [2]},\nonumber
\\
\hat{\mathcal F}^{z [9]}=-e^{\frac{2}{3}\phi}\big({\mathcal
F}^{[9]}-{\mathcal F}^{[8]}\wedge
\big(dz+A^{[1]}\big)\big),\nonumber
\\
\hat{\mathcal F}^{z [2]}=e^{-\frac{2}{3}\phi}\big({\mathcal
F}^{[2]}-{\mathcal F}^{[1]}\wedge \big(dz+A^{[1]}\big)\big),\qquad
{\mathcal F}^{[9]}=\ast {\mathcal F}^{[1]},\qquad {\mathcal
F}^{[8]}=e^{-\frac{3}{2}\phi}\ast {\mathcal F}^{[2]},
\end{gather*}
where ${\mathcal F}^{[8]}$, ${\mathcal F}^{[9]}$ were previously
def\/ined in \eqref{F8dr}, \eqref{F9dr} (they are not equal to
zero of course; the latter shall follow from the action).  With
the following ansatz for $\hat{v}$ \cite{bns}
\begin{gather*}
{\hat v}=e^{\frac{1}{12}\phi}v,\qquad i_v \big(dz+A^{[1]}\big)=0,
\qquad v=\frac{d a(x)}{{\sqrt{-\partial_{m}a\,
g^{mn}\,\partial_{n}a}}},\qquad v\ne v(z),
\end{gather*}
and using the standard rules of dimensional reduction
\cite{lpss95,popelec} (see also \cite{bns} for details of
reduction within the PST-type approach) one arrives at
\begin{gather*}
S_{\rm PST}=-\frac{1}{2} \int_{{\mathcal M}^{10}} v\wedge
{\mathcal F}^{a[8]}\wedge i_v {\mathcal F}^{b[2]}\eta_{ab}-v\wedge
{\mathcal F}^{[8]}\wedge i_v {\mathcal F}^{[2]}-v\wedge {\mathcal
F}^{[9]}\wedge i_v {\mathcal F}^{[1]}.
\end{gather*}
Hence the reduction of the complete duality-symmetric ${\rm D}=11$
gravity action $S=S_{\rm EHP}+S_{\rm PST}$ results in
\begin{gather}
S=-\int_{{\mathcal M}^{10}} \left(R^{ab}\wedge
\Sigma_{ab}+\frac{1}{2}v\wedge {\mathcal F}^{a[8]}\wedge i_v
{\mathcal
F}^{b[2]}\eta_{ab}\right)\nonumber\\
\phantom{S=}{}+\frac{1}{2} \int_{{\mathcal M}^{10}}\,
\sum_{n=1}^{2}\left(F^{[10-n]}\wedge F^{[n]}+i_v {\mathcal
F}^{[10-n]}\wedge v \wedge F^{[n]}+v\wedge F^{[10-n]}\wedge i_v
{\mathcal F}^{[n]}\right),\label{dsg10}
\end{gather}
where to recover this form of the action we have applied
identities like (\ref{idD3}) (see \cite{bns}).

In the latter form it is easy to recognize the duality-symmetric
structure of ${\rm D}=10$ gravity action, and the second line of
(\ref{dsg10}) is precisely the action \eqref{dsL10} we are looking
for. Therefore, we have established the relevance of the proposed
$S_{\rm PST}$ term in the action for the duality-symmetric
gravity.

One may wonder why having a new term entering the action we did
not modify $\hat{\tilde G}^{{\hat a}[2]}$ though~(\ref{dsgL11}) is
not a topological term and therefore it should contribute into the
energy-momentum tensor. The reason is the same as for another
application of the PST approach. Additional terms which follow
from the PST part of the action do not spoil the dynamics of an
original theory since they enter equations of motion only in
combinations of the generalized f\/ield strengths similar to
(\ref{F2a11}), (\ref{F9a11}) which are zero on-shell. The latter
is guaranteed by one of the PST symmetries
\cite{pst95plb,pst95prd,pst97prd}. That is for instance why we do
not need to take into account the contribution of two last terms
of \eqref{dsg10} into ${\mathfrak G}^{a[2]}$.

To give more rigorous arguments in favor of the discussion above,
let us consider a variation of the action
\begin{gather*}
S=S_{\rm EHP}+S_{\rm PST}
\end{gather*}
with $S_{\rm EHP}$ of \eqref{Dgrac} and $S_{\rm PST}$ of
\eqref{dsgL11}. The variation of the Einstein--Hilbert--Palatini
term results in
\begin{gather}\label{vEH}
\delta S_{\rm EHP}=\int_{{\mathcal M}^{11}} \delta {\hat e}^{\hat
a} \wedge d\hat{\mathcal F}^{{\hat b} [9]}\eta_{{\hat a}{\hat b}}.
\end{gather}
As for the PST term, it is convenient to split its variation into
the standard part for such an approach, where we will
ef\/fectively treat the vielbeins on the same footing with other
gauge f\/ields completely independent of metric, and the variation
over the metric tensor
\begin{gather*}
\delta S_{\rm PST}=\int_{{\mathcal M}^{11}} \delta_0 {\mathcal
L}_{\rm PST}+\delta_{*}{\mathcal L}_{\rm PST}.
\end{gather*}

The standard variation ends up with
\begin{gather}
\delta_0 S_{\rm PST}=\int_{{\mathcal M}^{11}} \left(\delta
\hat{\mathbb A}^{{\hat a}[8]}+\frac{\delta a}{{\sqrt{-(\partial
a)^2}}} i_{\hat v} \hat{\mathcal F}^{{\hat a}[8]}\right)
\eta_{{\hat a}{\hat b}} \wedge d\left(\hat{v}\wedge i_{\hat
v}\hat{\mathcal
F}^{{\hat b}[2]}\right)\notag\\
\phantom{\delta_0 S_{\rm PST}=}{}+\int_{{\mathcal M}^{11}}
\left(\delta \hat{e}^{\hat a} +\frac{\delta a}{{\sqrt{-(\partial
a)^2}}} i_{\hat v} \hat{\mathcal F}^{{\hat a}[2]}\right)
\eta_{{\hat a}{\hat b}} \wedge d\left(\hat{v}\wedge i_{\hat
v}\hat{\mathcal
F}^{{\hat b}[9]}\right)\notag\\
\phantom{\delta_0 S_{\rm PST}=}{}-\int_{{\mathcal M}^{11}}
\left(\delta \hat{e}^{\hat a} \eta_{{\hat a}{\hat b}} \wedge
d\hat{\mathcal F}^{{\hat b}[9]}+\delta
[\hat{\ast}\hat{\tilde{G}}^{{\hat a}[2]}] \eta_{{\hat a}{\hat b}}
\wedge \hat{v} \wedge i_{\hat v} \hat{\mathcal F}^{{\hat b}[2]}
\right).\label{v0PST}
\end{gather}

To deal with the other part of the variation we shall follow the
same way as upon deriving the energy-momentum tensor. In terms of dif\/ferential forms this general variation is
\begin{gather*}
\delta_* \big(\Omega^{[n]}\wedge \ast
\Omega^{[n]}\big)=\big(\delta_* \Omega^{[n]}\big)\wedge \ast
\Omega^{[n]}+\Omega^{[n]}\wedge \delta_*
\big(\ast \Omega^{[n]}\big)\notag\\
\qquad{}=\frac{1}{(n-1)!}\delta e^{a_n}\wedge
e^{a_{n-1}}\wedge\cdots\wedge e^{a_1} \Omega^{[n]}_{a_1 \cdots
a_{n-1} a_n}\wedge \ast
\Omega^{[n]}\notag\\
\qquad{} +\frac{1}{n!}\frac{1}{(D-n-1)!}\Omega^{[n]}\wedge \delta
e^{b_{D-n}}\wedge e^{b_{D-n-1}}\wedge\cdots\wedge
e^{b_1}{\epsilon_{b_1 \cdots b_{D-n}}}^{a_1 \cdots a_n}
\Omega^{[n]}_{a_1 \cdots a_n},
\end{gather*}
so applying it to our case leads to
\begin{gather}
\delta_* S_{\rm PST}=\int_{{\mathcal M}^{11}} \frac{1}{2} \hat{v}
\wedge \delta \hat{e}^{{\hat c}_9}\left(\frac{1}{8!}\hat{e}^{{\hat
c}_8}\wedge\cdots\wedge \hat{e}^{{\hat c}_1}{\hat{\mathcal
F}_{{\hat c}_1\cdots {\hat c}_8 {\hat c}_9}}^{~{\hat
a}[9]}\right)\eta_{{\hat a}{\hat b}}\wedge
i_{\hat v}\hat{\mathcal F}^{{\hat b}[2]}\notag\\
\phantom{\delta_* S_{\rm PST}=}{} +\int_{{\mathcal M}^{11}}
\frac{1}{2} \hat{v}\wedge \hat{\mathcal F}^{{\hat
a}[9]}\eta_{{\hat a}{\hat b}} \wedge \delta \hat{e}^{\hat c}
\left(i_{\hat v} \hat{\mathcal F}^{{\hat b}[2]}\right)_{\hat
c}.\label{v*PST}
\end{gather}
Taking into account \eqref{vEH}, \eqref{v0PST}, \eqref{v*PST}, one
can derive the following sets of the special symmetries of the
action
\begin{gather}\label{PST1}
\delta a(\hat{x})=0,\qquad \delta \hat{e}^{\hat a}=da\wedge
\hat{\varphi}^{{\hat a}[0]},\qquad \delta \hat{\mathbb A}^{{\hat
a}[8]}=da\wedge \hat{\varphi}^{{\hat
a}[7]}-d^{-1}\delta\left(\hat{\ast}\hat{\tilde{G}}^{{\hat
a}[2]}\right),
\\
\delta a(\hat{x})=\Phi (\hat{x}),\qquad \delta \hat{e}^{\hat
a}=-\frac{\Phi}{{\sqrt{-(\partial a)^2}}}\, i_{\hat v}
\hat{\mathcal
F}^{{\hat a}[2]},\notag\\
\delta \hat{\mathbb A}^{{\hat
a}[8]}=-\frac{\Phi}{{\sqrt{-(\partial a)^2}}} \,i_{\hat v}
\hat{\mathcal F}^{{\hat
a}[9]}\notag\\
\phantom{\delta \hat{\mathbb A}^{{\hat a}[8]}=}{}
-d^{-1}\left(\delta(\hat{\ast}\hat{\tilde{G}}^{{\hat
a}[2]})+\frac{1}{2!8!}\delta\hat{e}^{{\hat c}_9}\wedge
\hat{e}^{{\hat c}_8}\wedge\cdots\wedge \hat{e}^{{\hat
c}_1}\hat{\mathcal F}^{{\hat a}[9]}_{{\hat c}_1\dots {\hat
c}_8{\hat c}_9}-\frac{1}{2}\hat{\ast}\left(\delta \hat{e}^{\hat
c}\hat{\mathcal F}^{{\hat a}[2]}\right)_{\hat
c}\right).\label{PST2}
\end{gather}
with local gauge parameters $\hat{\varphi}^{\hat{a}[0]}$,
$\hat{\varphi}^{\hat{a}[7]}$ and $\Phi$.

The PST transformations of the ``graviton dual'' f\/ield are
non-local, but this non-locality does not spoil the job of the PST
symmetries \eqref{PST1}, \eqref{PST2} to extract the duality
relations and to establish pure auxiliary nature of the PST scalar
f\/ield.

To demonstrate that, it has to be noticed that the general
solution to the equation of motion of $\hat{\mathbb A}^{{\hat
a}[8]}$
\begin{gather*}
d\left(\hat{v}\wedge i_{\hat v}\hat{\mathcal F}^{{\hat
a}[2]}\right)=0
\end{gather*}
has the following form \cite{pst95plb,pst95prd,pst97prd}
\begin{gather}\label{solA8eom}
\hat{v}\wedge i_{\hat v}\hat{\mathcal F}^{{\hat a}[2]}=da\wedge
d\hat{\xi}^{{\hat a}[0]}.
\end{gather}
Upon the action of \eqref{PST1} the l.h.s.\ of the latter
expression is transformed as
\begin{gather*}
\hat{v}\wedge i_{\hat v}\hat{\mathcal F}^{{\hat
a}[2]}\longrightarrow \hat{v}\wedge i_{\hat v}\hat{\mathcal
F}^{{\hat a}[2]}+da\wedge d\hat{\varphi}^{{\hat a}[0]},
\end{gather*}
and therefore, setting $\hat{\xi}^{{\hat
a}[0]}=\hat{\varphi}^{{\hat a}[0]}$, one can reduce
\eqref{solA8eom} to
\begin{gather*}
i_{\hat v}\hat{\mathcal F}^{{\hat a}[2]}=0 \quad\rightsquigarrow
\quad \hat{\mathcal F}^{{\hat a}[2]}=0.
\end{gather*}
Taking it into account and using the same arguments one can reduce
the vielbein equation of motion to
\begin{gather*}
\hat{\mathcal F}^{{\hat a}[9]}=0.
\end{gather*}
The equation of motion of the PST scalar $a(\hat{x})$ is
identically satisf\/ied as a consequence of the equations of
motion for the other f\/ields. Therefore, the PST scalar equation
of motion is the Noether identity corresponding to an additional
symmetry which is nothing but the symmetry under the f\/irst of
the transformations \eqref{PST2}.

\subsection[Duality-symmetric action for the bosonic subsector of
${\rm D}=11$ supergravity]{Duality-symmetric action for the
bosonic subsector\\ of $\boldsymbol{{\rm D}=11}$ supergravity}

Let us now extend the results to the bosonic sector of ${\rm
D}=11$ supergravity. Note f\/irst that once a three-index photon
f\/ield $\hat{A}^{[3]}$ is taken into account equation \eqref{Fam}
is replaced with
\begin{gather*}
\hat{F}^{[9]}_{\hat a}=d\hat{\mathbb A}^{[8]}_{\hat
a}+\hat{\ast}\hat{\tilde{\mathbb G}}^{[2]}_{\hat a}\equiv
\hat{\mathbb F}^{[9]}_{\hat a}+\hat{\ast}\hat{\tilde{\mathbb
G}}^{[2]}_{\hat a},
\end{gather*}
where $\hat{\tilde{\mathbb G}}^{[2]}_{\hat a}$ is def\/ined by
$\hat{\ast}\hat{\tilde{\mathbb G}}^{[2]}_{\hat
a}=\hat{S}^{[D-2]}_{\hat a}+(-)^{\frac{D(D-5)}{2}}
\hat{\square}^{-1} \hat{\boldsymbol \delta}\hat{\ast}\hat{\mathbb
J}^{[1]}_{\hat a}$ with $\hat{\ast}\hat{\mathbb J}^{[1]}_{\hat
a}=\hat{\ast}\hat{J}^{[1]}_{\hat a}+\hat{M}^{[10]}_{\hat a}$. Here
as in equation \eqref{frJ} we have introduced the $\hat{A}^{[3]}$
energy-momentum form $\hat{M}^{[10]}_{\hat a}$. In ef\/fect, the
reduction ans\"atze \eqref{Fa9s}, \eqref{Fz9s} shall be modif\/ied
as follows
\begin{gather}
\hat{\mathbb F}^{a [9]}=e^{\frac{4}{3}\phi}\ast {\mathbf g}^{a
[1]}\notag \\
\phantom{\hat{\mathbb F}^{a [9]}=}{}+e^{-\frac{1}{12}\phi}
\left(d{\mathbb A}^{a [7]}+\ast\tilde{\mathfrak G}^{a
[2]}+\frac{1}{12} \ast (d\phi\wedge e^a)+e^{-\frac{1}{12}\phi}\ast
{\mathbb G}^{a [2]}\right)\wedge
\big(dz+A^{[1]}\big),\label{Fa9s11}
\\
\hat{\mathbb F}^{z
[9]}=-e^{\frac{2}{3}\phi}\left(dA^{[8]}-\frac{3}{4} F^{[8]}\wedge
A^{[1]}+\frac{1}{2}B^{[2]}\wedge dB^{[6]}-\frac{1}{4}
F^{[6]}\wedge A^{[3]}\right)\notag\\
\phantom{\hat{\mathbb F}^{z [9]}=}{}-e^{\frac{2}{3}\phi}
\left(1+\frac{2}{3}\right)\ast d\phi +e^{\frac{4}{3}
\phi}\ast {\mathbf g}^{z[1]}\notag\\
\phantom{\hat{\mathbb F}^{z
[9]}=}{}+e^{\frac{2}{3}\phi}\left(dA^{[7]}+F^{[6]}\wedge
B^{[2]}+B^{[2]}\wedge B^{[2]}\wedge
dA^{[3]}+e^{-\frac{5}{6}\phi}\ast {\mathbb G}^{z[2]}\right)\wedge
\big(dz+A^{[1]}\big).\!\!\!\label{Fz9s11}
\end{gather}
To arrive at \eqref{Fa9s11}, \eqref{Fz9s11} we have used the
reduction rule $\hat{\tilde{\mathbb G}}^{[2]}_{\hat a}={\mathbb
G}^{[2]}_{\hat a}+{\mathbf g}^{[1]}_{\hat a}\wedge (dz+A^{[1]})$,
$\ast\tilde{\mathfrak G}^{a [2]}$ is an analog of the form
def\/ined previously in \eqref{drvil1} and extended with the
appropriate contribution from the energy-momentum tensor of RR
3-form $A^{[3]}$ and of NS 2-form $B^{[2]}$. The latter forms come
from the reduction of $\hat{A}^{[3]}$. Finally, $F^{[6]}$ and
$F^{[8]}$ are the f\/ield strengths dual to
$F^{[4]}=dA^{[3]}-dB^{[2]}\wedge A^{[1]}$ and $F^{[2]}=dA^{[1]}$
\begin{gather*}
F^{[6]}=dA^{[5]}+A^{[3]}\wedge dB^{[2]}-B^{[2]}\wedge dA^{[3]},
\\
F^{[8]}=dA^{[7]}+F^{[6]}\wedge B^{[2]}+B^{[2]}\wedge B^{[2]}\wedge
dA^{[3]},
\end{gather*}
and $B^{[6]}$ is the dual to $B^{[2]}$ gauge f\/ield.

Let us end up with the action for the complete duality-symmetric
bosonic sector of ${\rm D}=11$ supergravity
(cf.~\cite{bbs98,bns}). The action is as follows
\begin{gather}
S=\int_{{\mathcal M}^{11}} \left[ \hat{R}^{{\hat a}{\hat b}}\wedge
\hat{\Sigma}_{{\hat a}{\hat
b}}+\frac{1}{2}\hat{F}^{[4]}\wedge\hat{\ast}\hat{F}^{[4]}-\frac{1}{3}
\hat{A}^{[3]}\wedge \hat{F}^{[4]} \wedge
\hat{F}^{[4]}\right]\notag\\
\phantom{S=}{}+\int_{{\mathcal M}^{11}} \frac{1}{2}\left[
\hat{v}\wedge \hat{\mathcal F}^{{\hat a}[9]}\wedge i_{\hat v}
\hat{\mathcal F}^{{\hat b}[2]}\eta_{{\hat a}{\hat
b}}-\hat{v}\wedge \hat{\mathcal F}^{[7]}\wedge i_{\hat v}
\hat{\mathcal F}^{[4]}\right],\label{ds11sg}
\end{gather}
with
\begin{gather*}
\hat{\mathcal F}^{{\hat a}[2]}=d\hat{e}^{{\hat
a}}-\hat{\ast}\left(d\hat{\mathbb A}^{{\hat
a}[8]}+\hat{\ast}\hat{\tilde {\mathbb G}}^{{\hat a}[2]}\right),
\\
\hat{\mathcal F}^{{\hat a}[9]}=-\hat{\ast}\hat{\mathcal F}^{{\hat
a}[2]},
\\
\hat{F}^{[4]}=d \hat{A}^{[3]},\qquad \hat{F}^{[7]}=d \hat
{A}^{[6]}+\hat{A}^{[3]}\wedge \hat{F}^{[4]},
\\
\hat{\mathcal
F}^{[4]}=\hat{F}^{[4]}-\hat{\ast}\hat{F}^{[7]},\qquad
\hat{\mathcal F}^{[7]}=-\hat{\ast}\hat{\mathcal F}^{[4]}.
\end{gather*}

After dimensional reduction with taking into account the ans\"atze
\eqref{Fa9s11}, \eqref{Fz9s11} together with the Kaluza--Klein
decomposition of gauge f\/ields
\begin{gather*}
\hat{A}^{[3]}=A^{[3]}-B^{[2]}\wedge dz,\notag \\
\hat{A}^{[6]}=B^{[6]}-A^{[5]}\wedge dz, \notag \\
\hat{F}^{[4]}=F^{[4]}+dB^{[2]}\wedge \big(dz+A^{[1]}\big),\notag \\
\hat{F}^{[7]}=dB^{[6]}+A^{[3]}\wedge dA^{[3]}-F^{[6]}\wedge
A^{[1]}+F^{[6]}\wedge \big(dz+A^{[1]}\big),
\end{gather*}
the action \eqref{ds11sg} produces the following action of the
completely duality-symmetric type IIA supergravity (cf.\ equation
(69) in~\cite{bns})
\begin{gather*}
S=-\int_{{\mathcal M}^{10}} \left(R^{ab}\wedge
\Sigma_{ab}+\frac{1}{2}v\wedge {\mathcal F}^{a[8]}\wedge i_v
{\mathcal
F}^{b[2]}\eta_{ab}\right)\\
\phantom{S=}{}+\frac{1}{2} \int_{{\mathcal M}^{10}}
\sum_{n=1}^{4}\left(\frac{1}{3^{[{n+\frac{1}{4}]}}}
F^{[10-n]}\wedge F^{[n]}+i_v {\mathcal F}^{[10-n]}\wedge v \wedge
F^{[n]}+v\wedge F^{[10-n]}\wedge i_v {\mathcal
F}^{[n]}\right).\notag
\end{gather*}
Here we have used the following set of the f\/ield strengths
\begin{gather*}
F^{[1]}=d\phi,\qquad F^{[9]}=dA^{[8]}-\frac{3}{4} F^{[8]}\wedge
A^{[1]}+\frac{1}{2} B^{[2]}\wedge dB^{[6]}-\frac{1}{4}
F^{[6]}\wedge A^{[3]},\notag \\
F^{[2]}=dA^{[1]},\qquad F^{[8]}=dA^{[7]}+F^{[6]}\wedge
B^{[2]}+B^{[2]}\wedge B^{[2]}\wedge dA^{[3]},\notag \\
F^{[3]}=dB^{[2]},\qquad F^{[7]}=dB^{[6]}+A^{[3]}\wedge
dA^{[3]}-F^{[6]}\wedge A^{[1]},\notag \\
F^{[4]}=dA^{[3]}-F^{[3]}\wedge A^{[1]},\qquad
F^{[6]}=dA^{[5]}+A^{[3]}\wedge F^{[3]}-B^{[2]}\wedge dA^{[3]},
\end{gather*}
the set of the generalized f\/ield strengths
\begin{gather*}
\mathcal{F}^{a[2]}=de^{a}+\ast \big(d\mathbb{A}^{a[7]}+\ast
\tilde{\mathfrak{G}}^{a[2]}\big),\qquad \mathcal{F}^{a[8]}=\ast
\mathcal{F}^{a[2]},\notag \\
\mathcal{F}^{[1]}=F^{[1]}+\ast F^{[9]},\qquad
\mathcal{F}^{[9]}=\ast \mathcal{F}^{[1]},\notag \\
\mathcal{F}^{[3]}=F^{[3]}+e^{-\phi} \ast F^{[7]},\qquad
\mathcal{F}^{[7]}=\ast \mathcal{F}^{[3]},\notag \\
\mathcal{F}^{[n]}=F^{[n]}+e^{\frac{(10-2n)}{4} \phi} \ast
F^{[10-n]}, \quad n=2,4,\notag \\
\mathcal{F}^{[n+5]}=F^{[n+5]}+e^{-\frac{n}{2}\phi}\ast
F^{[5-n]},\quad n=1,3,
\end{gather*}
and $[n+\frac{1}{4}]$ denotes the integral part of the number
$n+\frac{1}{4}$.

This f\/inalizes our proof of (\ref{ds11sg}) to be the complete
duality-symmetric action of ${\rm D}=11$ supergravity in the zero
fermion setting.

\section{On-shell dual gravity in linearized
approximation}

Let us now turn to the discussion on the linearized limit of
gravity and dualization of gravity in this case. Such a limit is
interesting for several reasons. For instance, the linearized
gravity can be reformulated in dual variables in the same way as
it could be done for an Abelian-like theory (see e.g.\
\cite{gopr98,gor98,nieto99,hull98,hull00,hull01,ahs04,fh05} and
references therein for various treatments of the problem). There
are also duality-invariant f\/irst order Lagrangians of ${\rm
D}=4$ linearized gravity~\cite{ht05} with \cite{jlr05} or without
a cosmological constant. The construction of these actions is
based on the approach of~\cite{dt76} which applies to the
linearized ADM formulation of General Relativity~\cite{adm62}.
However, the results of \cite{ht05} as well as other results on
the dual description of gravitation theory, do not go beyond the
linearized approximation \cite{ds05} that does not contradict the
summary made in \cite{dt76}. Though there is no rigorous proof,
the same is expected for ${\rm D}=4$ (a)dS linearized gravity
\cite{jlr05}, since just introducing a cosmological constant would
not improve the situation. A reason for the ``no-go'' theorem of
\cite{dt76} is easy to understand: to reach the manifestly dual
formulation in a spirit of \cite{dt76}, the  constraints have to
be resolved. It is hard to do for the original ADM constraints
with keeping locality in the end, while it turns out to be
possible after the linearization \cite{ht05}. The dual description
of gravity as a self-interacting theory is hampered by other
``no-go" theorems (see e.g.~\cite{bb03,bbh03}) which claim that
the dual description of spin-2 self-interacting theory is
inaccessible with keeping the locality of the theory. We avoid
these ``no-go" theorems by putting non-locality on the scene. In
ef\/fect, a consistency requirement for our approach is locality
of the dual gravity in the linearized approximation.

To move on such an approximation we have to expand vielbeins near
the f\/lat space limit
\begin{gather}\label{linv}
\hat{e}^{\hat a}=dX^{\hat m} u^{\hat a}_{\hat
m}+\hat{\mathcal{E}}^{\hat a},
\end{gather}
where $u^{\hat a}_{\hat m}$ is a constant matrix. After that we
should f\/ind a representation for equations of motion (equation
(\ref{Rem3}) in our case) which would be linear in f\/ields.
Taking into account (\ref{linv}), one gets
\begin{gather*}
\hat{\omega}^{\hat{a}\hat{b}}(\hat{e}) \cong \frac{1}{2}~ dX^{\hat
k} u^{\hat c}_{\hat k} \left[ u^{\hat m}_{\hat c}
u^{\hat{n}\hat{a}}\partial_{[\hat{m}}\hat{\mathcal{E}}^{\hat
b}_{\hat{n}]} -u^{\hat m}_{\hat c}
u^{\hat{n}\hat{b}}\partial_{[\hat{m}}\hat{\mathcal{E}}^{\hat
a}_{\hat{n}]}-u^{\hat{n}\hat{a}} u^{\hat{s}\hat{b}}
\partial_{[\hat{n}} \hat{\mathcal{E}}_{\hat{s}] \hat{c}} \right]+
\mathcal{O}\big(\hat{\mathcal{E}}^2\big),
\end{gather*}
that means
\begin{gather*}
\hat{J}^{[1]}_{\hat a}\cong \mathcal{O}
\big(\hat{\mathcal{E}}^2\big),
\end{gather*}
hence it does not enter the linearized equation of motion, and
\begin{gather*}
\hat{S}^{[D-2]}_{\hat a}=\hat{\ast} \left( dX^{\hat k} u^{\hat
b}_{\hat k} \wedge dX^{\hat l} u_{\hat{l}\hat{a}} \right)~ u^{\hat
m}_{\hat c} u^{\hat n}_{\hat b} \partial_{[\hat{m}}
\hat{\mathcal{E}}^{\hat c}_{\hat{n}]} + \mathcal{O}
(\hat{\mathcal{E}}^2) \equiv \hat{\mathbb{S}}^{[D-2]}_{\hat
a}+\mathcal{O} \big(\hat{\mathcal{E}}^2\big).
\end{gather*}
Then, equation of motion (\ref{Rem3}) becomes
\begin{gather}\label{lineom}
d\left(\hat{\ast} d \hat{\mathcal{E}}_{\hat a} -
\hat{\mathbb{S}}^{[D-2]}_{\hat a}\right)=0.
\end{gather}

A non-locality corresponding to a self-interacting part of the
whole action disappears in the linearized limit. Moreover, we can
write down (\ref{lineom}) in a form which will be manifestly
invariant under the local Lorentz rotations
\begin{gather*}
d \left( \hat{\ast} d \big(\hat{\mathcal{E}}^{\hat a} u^{\hat
m}_{\hat a}\big) - \big(\hat{\mathbb{S}}^{[D-2] \hat a} u^{\hat
m}_{\hat a}\big) \right)=0.
\end{gather*}
As a consequence, the Lorentz transformations of the dual f\/ield will be free of non-locality.
These observations lend credence
to the local formulation of action for the duality-symmetric
linearized gravity. Details on the action we postpone to a
forthcoming paper.

\section{Summary}

We would like to summarize with the following points. The main
refrain of this paper is non-locality which arises upon
dualization. We have tried to shed some light on non-locality in
duality rotations, actions and of\/f-shell gauge transformations,
and have demonstrated that depending on the approach we follow it
is possible to keep locality upon dualization. However, it is kind of exceptional,
 rather  than general cases, and we
have to conclude that non-locality shall be expected upon the
construction of duality-symmetric theories.

We have reviewed a part of dynamical realization of the
$\mathrm{E}_{11}$ conjecture. There are other approaches which are
tightly related to the conjecture, and which we would like to
mention in brief. Many important observations in favor of
$\mathrm{E}_{11}$ were made within the Cosmological Billiards in
String Cosmology~\cite{dt04,dt05} (and references therein), as
well as within the $\mathrm{E}_{10}$ approach to M-theory
\cite{dhn02}. Properties of imaginary roots of $\mathrm{E}_{10}$
in correspondence with branes, higher order corrections to
M-theory and orbifolds were studied in \cite{bgh04,bggh05,dn05}. A
correspondence between M(atrix)-theory and $\mathrm{E}_{11}$ was
reviewed in~\cite{c04}, and interesting observations on
orbifolding $\mathrm{E}_{11}$ were made in~\cite{m05}. It is also
worth mentioning the role of Borcherds superalgebras in
 M-theory~\cite{hjp02,hjp03} and a Mysterious duality between the U-duality
group of (dimensionally reduced) M-theory and classical symmetries
of del Pezzo surfaces~\cite{inv01}.

Finally, in ${\rm D}=4$ Yang--Mills theory there exists the true
duality-symmetric on-shell formulation, which is based on the loop
space approach to the YM and on a generalization of the Hodge
duality to non-Abelian f\/ields~\cite{cft96}. It is a very
intriguing problem to realize the same for General Relativity.

\appendix
\section{Notation and conventions}

Our notation and conventions which we use throughout the paper are
as follows. We reserve the hat symbol to distinguish
$\mathrm{D}$-dimensional quantities from $\mathrm{D-1}$
dimensional ones. Letters from the middle of the Latin alphabet
are used for the curved indices, while letters from the beginning
stand for the indices in tangent space. We use the same notation
for groups and corresponding algebras.

We choose the mostly minus signature $\eta_{ab}=\mathrm{diag}(+--
\cdots --)$. The antisymmetric $\mathrm{D}$ dimensional
Levi-Civita tensor $\epsilon^{a_1 \cdots a_D}$ is def\/ined by
\begin{gather*}
\epsilon^{01 \cdots (D-1)}=1,\qquad \epsilon_{01 \cdots
(D-1)}=(-)^{D-1},
\end{gather*}
so that
\begin{gather*}
\epsilon^{a_1 \cdots a_D} \epsilon_{a_1 \cdots a_D}=(-)^{D-1} D!.
\end{gather*}
In a curved space the latter becomes
\begin{gather*}
\epsilon^{m_1 \cdots m_D} \epsilon_{m_1 \cdots m_D}=\det{g}\, D!.
\end{gather*}

An arbitrary $n$-form has the following representation in
holonomic and non-holonomic basis
\begin{gather*}
\omega^{[n]}=\frac{1}{n!} dx^{m_n}\wedge \cdots \wedge dx^{m_1}\,
\omega^{[n]}_{m_1 \cdots m_n}=\frac{1}{n!} e^{a_n}\wedge \cdots
\wedge e^{a_1}\, \omega^{[n]}_{a_1 \cdots a_n},
\end{gather*}
and the exterior derivative acts from the right.

We def\/ine the Hodge star as
\begin{gather*}
\ast (e^{b_n}\wedge \cdots \wedge
e^{b_1})=\frac{\alpha_n}{(D-n)!}\, e^{a_{D-n}} \wedge \cdots
\wedge e^{a_1} {\epsilon_{a_1 \cdots a_{D-n}}}^{b_1 \cdots b_n},
\end{gather*}
that in turn implies
\begin{gather*}
\ast\big(dx^{k_n} \wedge \cdots \wedge
dx^{k_1}\big)=\frac{1}{(D-n)!} \frac{\alpha_n}{\sqrt{|g|}}\,
dx^{m_{D-n}} \wedge \cdots \wedge dx^{m_1} {\epsilon_{m_1 \cdots
m_{D-n}}}^{k_1 \cdots k_{D-n}}.
\end{gather*}
The coef\/f\/icients $\alpha_n$ can be f\/ixed to obey
\begin{gather*}
\alpha_n \cdot \alpha_{D-n}=(-)^{(D-n)n+(D-1)},
\end{gather*}
that provides the universal identity
\begin{gather*}
\ast^2=1.
\end{gather*}
In odd space-time dimensions all $\alpha_n$ are equal to one,
while in even dimensions we have a freedom in f\/ixing their
values. In such cases we will consider
\begin{gather*}
\begin{cases}
\alpha_n=1, & n< D/2;\\
\alpha_n=(-)^{(D-n)n+(D-1)}, & n> D/2;\\
\alpha^2_n=(-)^{n+(D-1)}, & n=D/2.
\end{cases}
\end{gather*}

Finally, the $\mathrm{D}$-dimensional volume form is as follows
\begin{gather*}
dx^{m_1} \wedge \cdots \wedge dx^{m_D}=d^D x\, \epsilon^{m_1
\cdots m_D},
\\
e^{a_1}\wedge \cdots \wedge e^{a_D}= d^D x\, \det e\,
\epsilon^{a_1 \cdots a_D},
\end{gather*}
and the contraction of a form with a one-form $v$ is def\/ined by
\begin{gather*}
i_v \omega^{[n]}=v^a i_a \omega^{[n]}=v^m i_m
\omega^{[n]},\notag\\
i_a \omega^{[n]}=\frac{1}{(n-1)!}\, e^{a_{n-1}} \wedge \cdots
\wedge
e^{a_1} \omega_{a a_1 \cdots a_{n-1}},\notag \\
i_m \omega^{[n]}=\frac{1}{(n-1)!}\, dx^{m_{n-1}} \wedge \cdots
\wedge
dx^{m_1} \omega_{m m_1 \cdots m_{n-1}}.
\end{gather*}

\section{Low-energy, strong coupling and high energy limits\\ of String
theory}

In this Appendix we brief\/ly recall why the low-energy limit
corresponds to sending the string tension to inf\/inity, why the
strong coupling regime comes once the tension sends to zero, and
what is the dif\/ference between the high energy limit and the
strong coupling limit.

Let us begin with the latter. In String theory there is a relation
between the string tension~$T$, a spin of a strings' mode $s$ and
its mass $M$:
\begin{gather}\label{MTs}
M^2 \sim T\cdot s.
\end{gather}
A set of these relations def\/ines the Regge trajectories of
particles in the string spectrum.

There are also two dif\/ferent scales: the Plank scale $l_P$ and
the string scale $l_s$ which are related to each other via a
string coupling constant $g_s$
\begin{gather*}
l_P^3=g_s \cdot l_s^3 .
\end{gather*}
The string tension is measured in the string scale units
\begin{gather*}
T\sim l_s^2.
\end{gather*}
Then, keeping the Plank scale f\/ixed and sending the tension to
zero, it becomes $g_s \rightarrow \infty$ that is the {\tt strong
coupling limit}. In this limit all the string modes become
massless due to~(\ref{MTs}).

Let us now turn to the {\tt low-energy limit}. In this limit $g_s
\rightarrow 0$, hence to keep $l_P$ we have to send $l_s
\rightarrow \infty$ that is $T \rightarrow \infty$. In such a
limit massive states are so heavy that they become ``frozen'', and
dynamics is managed by massless states.

Finally, in the {\tt high-energy limit} of String theory the Plank
energy sends to inf\/inity, viz. $l_P \rightarrow \infty$. Then
massive states become ultra-heavy, only light modes are dynamical,
hence $l_s \rightarrow 0$ and $g_s \rightarrow \infty$.

\subsection*{Acknowledgements}

I am very indebted to Vladimir Grigorievich Zima and Anatoly
Il'ich Pashnev, who inspired me for studying Supergravity and
String Theory, for their encouragement and many useful discussions
on various topics of strings.

I'm very grateful to Bernard Julia for a brief but very
informative discussion on dual models of gravity, and to Xavier
Bekaert, Ruben Mkrtchyan and Yuri Zinoviev for discussions on
duality-symmetric theories. Special thanks to Igor Bandos and
Dmitri Sorokin for pleasant collaboration, constant encouragement
and enlightening me via discussions and correspondence.

This work is supported in part by the Ukrainian SFFR Grant \#
F7/336-2001.

\LastPageEnding


\begin{thebibliography}{99}
\footnotesize

\bibitem{apps97}
Aganagic M., Park J., Popescu C., Schwarz J.H., World-volume
action of the M theory f\/ive-brane, {\it Nucl. Phys. B}, 1997,
V.496, 191--214; hep-th/9701166.

\bibitem{ahs04}
Ajith K.M., Harikumar E., Sivakumar M., Dual linearized gravity in
arbitrary dimensions, {\it Class. Quant. Grav.}, 2005, V.22,
5385--5396; hep-th/0411202.


\bibitem{pereira05}
Arcos H.I., Pereira J.G., Torsion gravity: a reappraisal, {\it
Internat. J. Modern Phys. D}, 2004, V.13, 2193--2240;
gr-qc/0501017.

\bibitem{adm62}
Arnowitt R., Deser S., Misner C.W., The dynamics of General
Relativity, in Gravitation: an Introduction to Current Research,
Editor L.~Witten, New York, Wiley, 1962, 227--264; gr-qc/0405109.

\bibitem{bbs98}
Bandos I., Berkovits N., Sorokin D., Duality-symmetric
eleven-dimensional supergravity and its coupling to M-branes, {\it
Nucl. Phys. B}, 1998, V.522, 214--233; hep-th/9711055.

\bibitem{blnpst97}
Bandos I., Lechner K., Nurmagambetov A., Pasti P., Sorokin D.,
Tonin M., Covariant action for super-f\/ive-brane of M theory,
{\it Phys. Rev. Lett.}, 1997, V.78, 4332--4335; hep-th/9701149.

\bibitem{bns}
Bandos I.A., Nurmagambetov A.J., Sorokin D.P., Various faces of
type IIA supergravity, {\it Nucl. Phys. B}, 2004, V.676, 189--228;
hep-th/0307153.

\bibitem{b04}
Banks T., Landskepticism: or why ef\/fective potentials don't
count string models, hep-th/0412129.

\bibitem{deserseminara}
Bautier K., Deser S., Henneaux M., Seminara D., No cosmological
${\rm D}=11$ supergravity, {\it Phys. Lett. B}, 1997, V.406,
49--53; hep-th/9704131.

\bibitem{bb03}
Bekaert X., Boulanger N., Massless spin-two f\/ield S-duality,
{\it Class. Quant. Grav.}, 2003, V.20, S417--S424; hep-th/0212131.

\bibitem{bbh03}
Bekaert X., Boulanger N., Henneaux M., Consistent deformations of
dual formulations of linearized gravity: a no-go result, {\it
Phys. Rev.~D}, 2003, V.67, 044010, 8 pages; hep-th/0210278.

\bibitem{bk97}
Bengtsson I., Kleppe A., On chiral p-forms, {\it Int. J. Mod.
Phys. A}, 1997, V.12, 3397--3412; hep-th/9609102.

\bibitem{bkorvp01}
Bergshoef\/f E., Kallosh R., Ortin T., Roest D., Van Proeyen A.,
New formulation of ${\rm D}=10$ supersymmetry and D8-O8 domain
walls, {\it Class. Quant. Grav.}, 2001, V.18, 3359--3382;
hep-th/0103233.

\bibitem{b97}
Berkovits N., Local actions with electric and magnetic sources,
{\it Phys. Lett. B}, 1997, V.395, 28--35; hep-th/9610134.

\bibitem{blau87}
Blau M., Killing vectors, constraints and conserved charges in
Kaluza--Klein theories, {\it Class. Quant. Grav.}, 1987, V.4,
1207--1221.

\bibitem{bo74}
Borisov A., Ogievetsky V., Theory of dynamical af\/f\/ine and
conformal symmetries as gravity theory, {\it Teoret. Mat. Fiz.},
1974, V.21, 329--342 (in Russian).

\bibitem{bch03}
Boulanger N., Cnockaert S., Henneaux M., A note on spin-s duality,
{\it JHEP}, 2003, V.0306, 060, 18 pages; hep-th/0306023.

\bibitem{bggh05}
Brown J., Ganguli S., Ganor O.J., Helfgott C., E10 orbifolds, {\it
JHEP}, 2005, V.0506, 057, 60 pages; hep-th/0409037.

\bibitem{bgh04}
Brown J., Ganor O.J., Helfgott C., M-theory and E10: billiards,
branes, and imaginary roots, {\it JHEP}, 2004, V.0408, 063, 68
pages; hep-th/0401053.

\bibitem{cft96}
Chan H.-M., Faridani J., Tsou S.T., Generalized dual symmetry for
non-Abelian Yang--Mills f\/ields, {\it Phys. Rev. D}, 1996, V.53,
7293--7305; hep-th/9512173.

\bibitem{c04}
Chaudhuri S., Hidden symmetry unmasked: matrix theory and E(11),
hep-th/0404235.

\bibitem{cd80}
Christensen S.M., Duf\/f M.J., Quantizing gravity with a
cosmological constant, {\it Nucl. Phys. B}, 1980, V.170, 480--506.

\bibitem{cj79}
Cremmer E., Julia B., The SO(8) supergravity, {\it Nucl. Phys. B},
1979, V.159, 141--212.

\bibitem{cjlp98I}
Cremmer E., Julia B., L\"u H., Pope C.N., Dualisation of dualities
I, {\it Nucl. Phys. B}, 1998, V. 523, 73--144; hep-th/9710119.

\bibitem{cjlp98II}
Cremmer E., Julia B., L\"u H., Pope C.N., Dualisation of dualities
II: Twisted self-duality of doubled f\/ields and superdualities,
{\it Nucl. Phys. B}, 1998, V. 535, 242--292; hep-th/9806106.

\bibitem{cjs78}
Cremmer E., Julia B., Scherk J., Supergravity theory in 11
dimensions, {\it Phys. Lett. B}, 1978, V.76, 409--412.

\bibitem{dt04}
Damour T., Cosmological singularities, billiards and Lorentzian
Kac--Moody algebras, gr-qc/0412105.

\bibitem{dt05}
Damour T., Cosmological singularities, Einstein billiards and
Lorentzian Kac--Moody algebras, gr-qc/0501064.

\bibitem{dhn02}
Damour T., Henneaux M., Nicolai H., E10 and a ``small tension
expansion'' of M theory, {\it Phys. Rev. Lett.}, 2002, V.89,
221601, 4 pages; hep-th/0207267.

\bibitem{dn05}
Damour T., Nicolai H., Higher order M theory corrections and the
Kac--Moody algebra E10, {\it Class. Quant. Grav.}, 2005, V.22,
2849--2880; hep-th/0504153.

\bibitem{abp05}
de Andrade V.C., Barbosa A.L., Pereira J.G., Gravitation and
duality symmetry, gr-qc/0501037.

\bibitem{dd79}
Deser S., Drechsler W., Generalized gauge field copies, {\it Phys.
Lett. B}, 1979, V.86, 189--192.

\bibitem{dg98}
Deser S., Gibbons G.W., Born--Infeld--Einstein actions?, {\it Class.
Quant. Grav.}, 1998, V.15, L35--L39; hep-th/9803049.

\bibitem{djh84}
Deser S., Jackiw R., t'Hooft G., Three-dimensional Einsten
gravity: dynamics of f\/lat space, {\it Ann. Phys.}, 1984, V.152,
220--235.

\bibitem{dj84}
Deser S., Jackiw R., Three-dimensional cosmological gravity:
dynamics of constant curvature, {\it Ann. Phys.}, 1984, V.153,
405--416.

\bibitem{dj82}
Deser S., Jackiw R., Topologically massive gauge theories, {\it
Ann. Phys.}, 1982, V.140, 372--411.

\bibitem{dn82}
Deser S., Nepomechie R.I., Electric-magnetic duality of conformal
gravitation, {\it Phys. Lett. A}, 1983, V.97, 329--332.

\bibitem{ds05}
Deser S., Seminara D., Free spin 2 duality invariance cannot be
extended to general relativity, {\it Phys. Rev. D}, 2005, V.71, 081502, 7 pages;
hep-th/0503030.

\bibitem{dt76}
Deser S., Teitelboim C., Duality transformations of Abelian and
non-Abelian gauge f\/ields, {\it Phys. Rev. D}, 1976, V.13,
1592--1597.

\bibitem{dw76}
Deser S., Wilczek F., Non-uniqueness of gauge-field potentials,
{\it Phys. Lett. B}, 1976, V.65, 391--393.

\bibitem{wn00}
de Wit B., Nicolai H., Hidden symmetries, central charges and all
that, {\it Class. Quant. Grav.}, 2001, V.18, 3095--3112;
hep-th/0011239.

\bibitem{wn85}
de Wit B., Nicolai H., Hidden symmetries in ${\rm D}=11$
supergravity, {\it Phys. Lett.~B}, 1985, V.155, 47--53.

\bibitem{wn86}
de Wit B., Nicolai H., ${\rm D}=11$ supergravity with local SU(8)
invariance, {\it Nucl. Phys. B}, 1986, V.274, 363--400.

\bibitem{d04}
Douglas M.R., Basic results in vacuum statictics, {\it Comptes
Rendus Physique}, 2004, V.5, 965--977; hep-th/0409207.

\bibitem{gfs91}
Fradkin E.S., Gitman D.M., Shvartsman S.M., Quantum
electrodynamics with unstable vacuum, Berlin, Springer, 1991.

\bibitem{ft85}
Fradkin E.S., Tseytlin A.A., Quantum equivalence of dual f\/ield
theories, {\it Ann. Phys.}, 1985, V.162, 31--48.

\bibitem{fh05}
Francia D., Hull C.M., Higher-spin gauge f\/ields and duality,
hep-th/0501236.

\bibitem{fv71}
Fubini S., Veneziano G., Algebraic treatment of subsidiary
conditions in dual resonance model, {\it Ann. Phys.}, 1971, V.63,
12--27.

\bibitem{gow02}
Gaberdiel M.R., Olive D.I., West P.C., A class of Lorentzian
Kac--Moody algebras, {\it Nucl. Phys. B}, 2002, V.645, 403--437;
hep-th/0205068.

\bibitem{gopr98}
Garcia-Compean H., Obregon O., Plebansky J.F., Ramirez C., Towards
a gravitational analogue to S-duality in non-Abelian gauge
theories, {\it Phys. Rev. D}, 1998, V.57, 7501--7506;
hep-th/9711115.

\bibitem{gor98}
Garcia-Compean H., Obregon O., Ramirez C., Gravitational duality
in MacDowell--Mansouri gauge theory, {\it Phys. Rev. D}, 1998,
V.58, 104012, 3 pages; hep-th/9802063.

\bibitem{gsw87}
Green M.B., Schwarz J.H., Witten E., Superstring theory,
Cambridge, Cambridge University Press, 1987.

\bibitem{gross88}
Gross D.J., High-energy symmetries of string theory, {\it Phys.
Rev. Lett.}, 1988, V.60, 1229--1232.

\bibitem{ht05}
Henneaux M., Teitelboim C., Duality in linearized gravity, {\it
Phys. Rev. D}, 2005, V.71, 024018, 8 pages; gr-qc/0408101.

\bibitem{hjp02}
Henry-Labordere P., Julia B., Paulot L., Borcherds symmetries in
M-theory, {\it JHEP}, 2002, V.0204, 049, 31 pages; hep-th/0212346.

\bibitem{hjp03}
Henry-Labordere P., Julia B., Paulot L., Real Borcherds
superalgebras and M-theory, {\it JHEP}, 2003, V.0304, 060, 21 pages; hep-th/0203070.

\bibitem{hull98}
Hull C.M., Gravitational duality, branes and charges, {\it Nucl.
Phys. B}, 1998, V.509, 216--251; hep-th/9705162.

\bibitem{hull00}
Hull C.M., Strongly coupled gravity and duality, {\it Nucl. Phys.
B}, 2000, V.583, 237--259; hep-th/0004195.

\bibitem{hull01}
Hull C.M., Duality in gravity and higher spin gauge f\/ields, {\it
JHEP}, 2001, V.0109, 027, 25 pages; hep-th/0107149.


\bibitem{ht95}
Hull C.M., Townsend P.K., Unity of superstring dualities, {\it
Nucl. Phys. B}, 1995, V.438, 109--137; hep-th/9410167.

\bibitem{inv01}
Iqbal A., Neitzke A., Vafa C., A mysterious duality, {\it Adv.
Theor. Math. Phys.}, 2002, V.5, 769--808; hep-th/0111068.

\bibitem{julia82}
Julia B., Kac--Moody symmetry of gravitational and supergravity
theories, {\it Lect. Appl. Math.}, 1985, V.21, 355--375.

\bibitem{julia84}
Julia B., Supergeometry and Kac--Moody algebras, in Vertex
Operators in Mathematics and Physics,
New York, Springer,  1985, 393-410

\bibitem{jlr05}
Julia B., Levie J., Ray S., Gravitational duality near de Sitter
space, {\it JHEP}, 2005, V.0511, 025, 13 pages; hep-th/0507262.

\bibitem{k83}
Kac V., Inf\/inite dimensional Lie algebras, Birkhauser, Boston,
1983.

\bibitem{k99}
Kaku M., Introduction to superstrings and M-theory, 2nd ed.,
Berlin, Springer, 1999.

\bibitem{kns00}
Koepsell K., Nicolai H., Samtleben H., An exceptional geometry for
$d=11$ supergravity?, {\it Class. Quant. Grav.}, 2000, V.17,
3689--3702; hep-th/0006034.

\bibitem{linde05}
Linde A., Inf\/lation and string cosmology, {\it J. Phys. Conf.
Ser.}, 2005, V.24, 151--160; hep-th/0503195.

\bibitem{lpss95}
L\"u H., Pope C.N., Sezgin E., Stelle K.S., Stainless super
p-branes, {\it Nucl. Phys. B}, 1995, V.456, 669--698;
hep-th/9508042.

\bibitem{ms82}
Marcus N., Schwarz J.H., Field theories that have no manifestly
Lorentz-invariant formulation, {\it Phys. Lett.~B}, 1982, V.115,
111--114.

\bibitem{mr94}
Martin I., Restuccia A., Duality symmetric actions and canonical
quantization, {\it Phys. Lett. B}, 1994, V.323, 311--315.

\bibitem{msp99}
Maznytsia A., Preitschopf C.R., Sorokin D.P., Duality of self-dual
actions, {\it Nucl. Phys. B}, 1999, V.539, 438--452;
hep-th/9805110.

\bibitem{myw90}
McClain B., Yu F., Wu Y.S., Covariant quantization of chiral
bosons and $\mathrm{Osp}(1 \vert 2)$ symmetry, {\it Nucl. Phys.
B}, 1990, V.343, 689--704.

\bibitem{miz98}
Mizoguchi S., $E_{10}$ Symmetry in one-dimensional supergravity,
{\it Nucl. Phys. B}, 1998, V.528, 238--264; hep-th/9703160.

\bibitem{m05}
Mkrtchyan H., Mkrtchyan R., Remarks on E11 approach,
hep-th/0507183.

\bibitem{moore93}
Moore G., Finite in all directions, hep-th/9305139.

\bibitem{n87}
Nicolai H., ${\rm D}=11$ supergravity with local SO(16)
invariance, {\it Phys. Lett.~B}, 1987, V.187, 316--320.

\bibitem{nicolai92}
Nicolai H., A hyperbolic Lie algebra from supergravity, {\it Phys.
Lett. B}, 1992, V.276, 333--340.

\bibitem{npz05}
Nicolai H., Peeters K., Zamaklar M., Loop quantum gravity: an
outside review, {\it Class. Quant. Grav.}, 2005, V.22, R193--R247; hep-th/0501114.

\bibitem{nw89}
Nicolai H., Warner N.P., The structure of $N=16$ supergravity in
two dimensions, {\it Comm. Math. Phys.}, 1989, V.125, 369--384.

\bibitem{nieto99}
Nieto J.A., S-duality for linearized gravity, {\it Phys. Lett.~A},
1999, V.262, 274--281; hep-th/9910049.

\bibitem{ajnjetp04}
Nurmagambetov A.J., On the sigma-model structure of type IIA
supergravity action in doubled f\/ield approach, {\it JETP Lett.},
2004, V.79, 191--195; hep-th/0403100.

\bibitem{ajn04}
Nurmagambetov A.J., Duality-symmetric gravity and supergravity:
testing the PST approach, hep-th/0407116.

\bibitem{op99}
Obers N.A., Pioline B., U-duality and M-theory, {\it Phys. Rep.},
1999, V.318, 113--225; hep-th/9809039.

\bibitem{o73}
Ogievetsky V.I., Inf\/inite-dimensional algebra of general
covariance group as the closure of f\/inite dimensional algebras
of conformal and linear groups, {\it Lett. Nuov. Cim.}, 1973, V.8,
988--990.

\bibitem{pst95plb}
Pasti P., Sorokin D.P., Tonin M., Note on manifest Lorentz and
general coordinate invariance in duality symmetric models, {\it
Phys. Lett. B}, 1995, V.352, 59--63; hep-th/9503182.

\bibitem{pst95prd}
Pasti P., Sorokin D.P., Tonin M., Duality symmetric actions with
manifest space-time symmetries, {\it Phys. Rev. D}, 1995, V.52,
4277--4281; hep-th/9506109.

\bibitem{pst97prd}
Pasti P., Sorokin D.P., Tonin M., On Lorentz invariant actions for
chiral p-forms, {\it Phys. Rev. D}, 1997, V.55, 6292--6298;
hep-th/9611100.

\bibitem{peldan}
Peldan P., Actions for gravity, with generalizations: a review,
{\it Class. Quant. Grav.}, 1994, V.11, 1087--1132; gr-qc/9305011.

\bibitem{p98}
Polchinski J., String theory, Cambridge, Cambridge University
Press, 1998.

\bibitem{popelec}
Pope C.N., Lectures on Kaluza--Klein,
http://faculty.physics.tamu.edu/pope/

\bibitem{ss}
Salam A., Sezgin E., Supergravities in diverse dimensions, Vol.~1,
 Singapore, World Scientif\/ic, 1989.

\bibitem{sw01}
Schnakenburg I., West P., Kac--Moody symmetries of IIB
supergravity, {\it Phys. Lett. B}, 2001, V.517, 421--428;
hep-th/0107181.

\bibitem{sw02}
Schnakenburg I., West P., Massive IIA supergravity as a non-linear
realisation, {\it Phys. Lett. B}, 2002, V.540, 137--145;
hep-th/0204207.

\bibitem{ss94}
Schwarz J.H., Sen A., Duality symmetric actions, {\it Nucl. Phys.
B}, 1994, V.411, 35--63; hep-th/9304154.

\bibitem{s02}
Sorokin D., Lagrangian description of duality-symmetric f\/ields,
in Advances in the Interplay between Quantum and Gravity Physics,
Editors P.G.~Bergmann and V.~de Sabbata, Kluwer Academic Pub.,
2002, 365--385.

\bibitem{schro}
Schr\"odinger E., Space-time structure, Cambridge Univ. Press,
1950.

\bibitem{s03}
Susskind L., The anthropic landscape of string theory,
hep-th/0302219.

\bibitem{thirring86}
Thirring W., A course on mathematical physics, Vol.~2, Field
theory, 2nd ed., Berlin, Springer, 1986.

\bibitem{towns95}
Townsend P.K., The eleven-dimensional supermembrane revisited,
{\it Phys. Lett. B}, 1995, V.350, 184--187; hep-th/9501068.



\bibitem{vir70}
Virasoro M.A., Subsidiary conditions and ghosts in dual-resonance
model, {\it Phys. Rev. D}, 1970, V.1, 2933--2936.


\bibitem{w00}
West P., Hidden superconformal symmetry in M theory, {\it JHEP},
2000, V.0008, 007, 21 pages; hep-th/0005270.

\bibitem{w01}
West P., $\mathrm{E}_{11}$ and M theory, {\it Class. Quant.
Grav.}, 2001, V.18, 4443--4460; hep-th/0104081.

\bibitem{w02}
West P., Very extended $\mathrm{E}_8$ and $\mathrm{A}_8$ at low
levels, gravity and supergravity, {\it Class. Quant. Grav.}, 2002,
V.20, 2393--2406; hep-th/0212291.

\bibitem{w88}
Witten E., 2+1 dimensional gravity as an exactly soluble system,
{\it Nucl. Phys. B}, 1988/89, V.311, 46--78.

\bibitem{witten95}
Witten E., String theory dynamics in various dimensions, {\it
Nucl. Phys. B}, 1995, V.443, 85--126; hep-th/9503124.

\bibitem{w91}
Wotzasek C., The Wess--Zumino term for chiral bosons, {\it Phys.
Rev. Lett.}, 1991, V.66, 129--132.

\bibitem{wy75}
Wu T.T., Yang C.N., Some remarks about unquantized non-Abelian
gauge f\/ields, {\it Phys. Rev.~D}, 1975, V.12, 3843--3844.

\bibitem{z71}
Zwanziger D., Local Lagrangian quantum f\/ield theory of electric
and magnetic charges, {\it Phys. Rev.~D}, 1971, V.3, 880--890.



\end{thebibliography}
\end{document}